\documentclass[11pt]{article}

\usepackage{graphicx, color, url}
\usepackage{amsmath}
\usepackage{dsfont}
\usepackage{amssymb}
\usepackage{epstopdf}
\usepackage{boxedminipage}
\usepackage{amsfonts}
\usepackage{url}
\usepackage{amsbsy}
\usepackage{setspace}
\usepackage{natbib}
\usepackage{bm}
\usepackage{wasysym}
\usepackage{textcomp}

\usepackage {graphicx}
\usepackage{amssymb}
\usepackage{amsmath}
\usepackage{amsbsy}
\usepackage{setspace}
\usepackage{natbib}
\usepackage{bm}
\usepackage{mathtools}

\usepackage{wasysym}
\usepackage{textcomp}
\usepackage{amsmath,amssymb,amsfonts, amsbsy, epsfig, subfig,natbib,soul,hyperref}
\usepackage{algpseudocode}
\usepackage{algorithm}

\newtheorem{theorem}{Theorem}
\newtheorem{corollary}{Corollary}
\newtheorem{proposition}{Proposition}
\newtheorem{lemma}{Lemma}
\newtheorem{example}{Example}

\newtheorem{remark}{Remark}
\newtheorem{definition}{Definition}
\newtheorem{proc}{Procedure}

\newcommand{\bdelta}{ \mbox{\boldmath $\delta$}}

\newcommand{\beq}{\begin{equation}}
\newcommand{\eeq}{\end{equation}}
\newcommand{\beas}{\begin{eqnarray*}}
	\newcommand{\eeas}{\end{eqnarray*}}
\newcommand{\bea}{\begin{eqnarray}}
\newcommand{\eea}{\end{eqnarray}}
\newcommand{\bei}{\begin{itemize}}
	\newcommand{\eei}{\end{itemize}}
\newcommand{\ben}{\begin{enumerate}}
	\newcommand{\een}{\end{enumerate}}
\newcommand{\bet}{\begin{theorem}}
	\newcommand{\eet}{\end{theorem}}
\newcommand{\bel}{\begin{lemma}}
	\newcommand{\eel}{\end{lemma}}
\newcommand{\bep}{\begin{proposition}}
	\newcommand{\eep}{\end{proposition}}
\newcommand{\bed}{\begin{definition}}
	\newcommand{\eed}{\end{definition}}
\newcommand{\bec}{\begin{corollary}}
	\newcommand{\eec}{\end{corollary}}
\newcommand{\bex}{\begin{example}}
	\newcommand{\eex}{\end{example}}

\newcommand{\EE}{\mathbb{E}}

\def\0{\boldsymbol{0}}

\newcommand{\hQ}{\widehat{Q}}
\newcommand{\hTor}{\widehat{T}_{OR}}
\newcommand{\hU}{\widehat{U}}
\newcommand{\Tor}{T_{OR}}

\newcommand{\fne}{\tilde{f}_{1, \sigma}}
\newcommand{\fem}{\hat{f}_{1,\sigma}}
\newcommand{\fems}{\hat{f}^*_{1,\sigma}}
\newcommand{\sumind}{\sum_{i=1}^{m}\phi_{h_\sigma}(\sigma-\sigma_i)\mathbb{I}(\theta_i=1)}
\newcommand{\sumprob}{\sum_{i=1}^{m}\phi_{h_\sigma}(\sigma-\sigma_i)P(\theta_i=1|x_i,\sigma_i)}
\newcommand{\Et}{\mathbb{E}_{\pmb{\theta|\sigma,x}}}
\usepackage[svgnames]{xcolor}
\usepackage{listings}

\begin{document}
	
	\title{{Heterocedasticity-Adjusted Ranking and Thresholding for Large-Scale Multiple Testing}}
	
	\author{Luella Fu$^{1}$, \ Bowen Gang$^{2}$, \ Gareth M. James$^{3}$,\ and \ Wenguang Sun$^{3}$}
	
	\date{}
	
	\maketitle
	
	\begin{abstract}
		
	Standardization has been a widely adopted practice in multiple testing, for it takes into account the variability in sampling and makes the test statistics comparable across different study units. However, { despite conventional wisdom to the contrary, we show that} there can be a significant loss in information from basing hypothesis tests on standardized statistics rather than the full data. We develop a new class of heteroscedasticity--adjusted ranking and thresholding (HART) rules that aim to improve existing methods by simultaneously exploiting commonalities and adjusting heterogeneities among the study units. The main idea of HART is to bypass standardization by directly incorporating both the summary statistic and its variance into the testing procedure. A key message is that the variance structure of the alternative distribution, which is subsumed under standardized statistics, is highly informative and can be exploited to achieve higher power. The proposed HART procedure is shown to be asymptotically valid and optimal for false discovery rate (FDR) control. Our simulation {results} demonstrate that HART achieves substantial power gain over existing methods at the same FDR level. We illustrate the implementation through a microarray analysis of myeloma.

	\end{abstract}

	
	\footnotetext[1]{Department of Mathematics, San Francisco State University. }
	
	\footnotetext[2]{Department of Mathematics, University of Southern California. }
	
	\footnotetext[3]{Department of Data Sciences and Operations, University of Southern California. }
	

	\newpage
	
	
	\section{Introduction}\label{sec:intro}

In a wide range of modern scientific studies, multiple testing frameworks have been routinely employed by scientists and researchers to identify interesting cases among thousands or even millions of features. A representative sampling of settings where multiple testing has been used includes: genetics, for the analysis of gene expression levels \citep{Tushetal2001, Dudetal03, SunWei11}; astronomy, for the detection of galaxies \citep{Mil01}; neuro-imaging, for the discovery of differential brain activity \citep{Pacetal04, Schetal08}; education, to identify student achievement gaps \citep{Efr08b}; data visualization, to find potentially interesting patterns \citep{Zhao17}; and finance, to evaluate trading strategies \citep{HarLiu15}.

The standard practice involves three steps: reduce the data in different study units to a vector of summary statistics; standardize the summary statistics to obtain significance indices such as $z$-values or $p$-values; and  find a threshold of significance that corrects for multiplicity.
%
Given a summary statistic $X_i$ with associated standard deviation $\sigma_i$, traditional multiple testing approaches begin by standardizing the observed data $Z_i=X_i/\sigma_i$, which is then used to compute the $p$-value based on a problem specific null distribution.
Finally, the $p$-values are ordered, and a threshold is applied to 
keep the 
rate of Type I error below a pre-specified level.

Classical approaches concentrated on setting a threshold that controls the family-wise error rate (FWER), using methods such as the Bonferroni correction or Holm's procedure \citep{Hol79}. However, the FWER criterion becomes infeasible once the number of hypotheses under consideration grows to thousands. The seminal contribution of \citet{BenHoc95} proposed replacing the FWER by the false discovery rate (FDR) and provided the BH algorithm for choosing a threshold on the ordered $p$-values which, under certain assumptions, is guaranteed to control the FDR.

While the BH procedure offers a significant improvement over classical approaches, it only controls the FDR at level $(1-\pi)\alpha$, where $\pi$ is the proportion of non-nulls, suggesting that its power can be improved by incorporating an adjustement for $\pi$ into the procedure. \citet{BenHoc00}, \citet{Sto02} and \cite{GenWas02} proposed to first estimate the non-null proportion by $\hat\pi$ and then run BH at level $\alpha/(1-\hat\pi)$. \citet{Efretal01} proposed the local false discovery rate (Lfdr), which incorporates, in addition to the sparsity parameter $\pi$, information about the alternative distribution.
\citet{SunCai07} proved that the $z$-value optimal procedure is an Lfdr thresholding rule and that this rule uniformly dominates the $p$-value optimal procedure in \cite{GenWas02}.
The key idea is that the shape of the alternative could potentially affect the rejection region but the important structural information is lost when converting the $z$-value to $p$-value. For example, when the means of non-null effects are more likely to be positive than negative, then taking this asymmetry of the alternative into account increases the power. However, the sign information is not captured by conventional $p$-value methods, which only consider information about the null. 

Although a wide variety of multiple testing approaches have been proposed, they almost all begin with the standardized data $Z_i$ (or its associated $p$-value, $P_i$). In fact, in large-scale studies where the data are collected from intrinsically diverse sources, the standardization step has been upheld as conventional wisdom, for it takes into account the variability of the summary statistics and suppresses the heterogeneity -- enabling one to compare multiple study units on an equal footing. For example, in microarray studies, \cite{Efretal01} first compute standardized two-sample $t$-statistics for comparing the gene expression levels across two biological conditions and then convert the $t$-statistics to $z$--scores, which are further employed to carry out FDR analyses. Binomial data is also routinely standardized by rescaling the number of successes $X_i$ by the number of trials $n_i$ to obtain success probabilities $\hat p_i=X_i/n_i$ and then converting the probabilities to $z$-scores 
\citep{Efr08b, Efr08a}.
However, while standardization is an intuitive, and widely adopted, approach, we argue in this paper that there can be a significant loss in information from basing hypothesis tests on $Z_i$ rather then the full data $(X_i, \sigma_i)$\footnote{\small Unless otherwise stated, the term ``full data'' specifically refers to the pair $(X_i, \sigma_i)$ in this article. In practice, the process of deriving the pair $(X_i, \sigma_i)$ from the original (full) data could also suffer from information loss, but this point is beyond the scope of this work; see the rejoinder of Cai et al. (2019) for related discussions.}. This observation, which we formalize later in the paper, is based on the fact that the power of tests can vary significantly as $\sigma$ changes, but this difference in power is suppressed when the data is standardized and treated as equivalent. In the illustrative example in Section \ref{example:sec}, we show that by accounting for differences in $\sigma$ an alternative ordering of rejections can be obtained, allowing one to identify more true positives at the same FDR level.

This article develops a new class of heteroscedasticity-adjusted ranking and thresholding (HART) rules for large-scale multiple testing that aim to improve existing methods by simultaneously exploiting commonalities and adjusting heterogeneities among the study units. The main strategy of HART is to bypass standardization by directly incorporating $(X_i, \sigma_i)$ into the testing procedure. We adopt a two-step approach. In the first step a new significance index is developed by taking into account the alternative distribution of each $X_i$ conditioned on $\sigma_i$; hence HART avoids power distortion. This kind of conditioning is not possible for standardized values since the $\sigma_i$ are subsumed under $Z_i$. Then, in the second step the significance indices are ordered and the smallest ones are rejected up to a given cutoff. We develop theories to show that HART is optimal for integrating the information from both $X_i$ and $\sigma_i$. Numerical results are provided to confirm that HART controls the FDR in finite samples and uniformly dominates existing methods in power.

{ We are not the first to consider adjusting for heterogeneity.} {
\cite{Ignetal16} and \cite{LeiFit18} mentioned the possibility of using the $p$-value as a primary significance index  while employing $\sigma_i$ as side-information to pre-order hypotheses. Earlier works by \cite{Efr08b} and \cite{CaiSun09} also suggest grouping methods to adjust for heterogeneous variances in data. However, the variance issue is only briefly mentioned in these works and it is unclear how a proper pre--ordering or grouping can be created based on $\sigma_i$. It is important to note that the ordering or grouping based on the magnitudes of $\sigma_i$ will not always be informative. Concretely, a large $\sigma_i$ does not generally correspond to a hypothesis that is more or less likely to be a true signal. Our numerical results suggest that ordering by $\sigma_i$ can create a somehow arbitrary ordering of hypotheses, which can even be anti-informative, potentially leading to power loss compared to methods that utilize no side information. In contrast with existing works, we {explicitly demonstrate the key role that $\sigma_i$ plays} in characterizing the shape of the  alternative in simultaneous testing (Section \ref{example:sec}). Moreover, we develop a principled and optimal strategy, the HART procedure, for incorporating the structural information encoded in $\sigma_i$ into inference. We prove that HART guarantees FDR control and uniformly improves upon all existing methods in asymptotic power. }

The findings are impactful for three reasons. First, the observation that standardization can be inefficient 
has broad implications since, due to inherent variabilities or differing sample sizes between study units, standardized  tests are commonly applied to large-scale heterogeneous data to make different study units comparable. Second, our finding enriches the recent line of research on multiple testing with side and structural information (e.g. \citealp{Caietal19, LiBar19, Xiaetal19}, among others). In contrast with these works that have focused on the usefulness of sparsity structure, our characterization of the impact of heteroscedasticity, or more concretely \emph{the shape of alternative distribution}, is new. Finally, HART convincingly demonstrates the benefits of leveraging structural information in high-dimensional settings when the number of tests is in the thousands or more. Ideas from HART apply to smaller data sets as well, but the algorithm is designed to capitalize on copious data in ways not possible for procedures intended for moderate amounts of data, and thus is most useful in large-scale testing scenarios where the structure can be learned from data with good precision.

The rest of the paper is organized as follows. Section~\ref{sec:or} reviews the standard multiple testing model and provides a motivating example that clearly illustrates the potential power loss from standardization. Section~\ref{sec:method} describes our HART procedure and its theoretical properties.  Section \ref{sec:simu} contains simulations, and Section \ref{sec:data} demonstrates the method on a microarray study. We conclude the article with a discussion of connections to existing work and open problems. Technical materials and proofs are provided in the Appendix.

	\section{Problem Formulation and the Issue of Standardizing}\label{sec:or}

\setcounter{equation}{0}

This section first describes the problem formulation and then discusses an example to illustrate the key issue.

\subsection{Problem formulation}\label{prob-form:sec}

{Let $\theta_i$ denote a Bernoulli($\pi$) variable, where $\theta_i=0/1$ indicates a null/alternative hypothesis, and $\pi = P(\theta_i=1)$ is the proportion of nonzero signals coming from the alternative distribution. Suppose the summary statistics $X_1, \ldots, X_m$ are normal variables obeying distribution 
\begin{equation}\label{x.eqn}
X_i|\mu_i, \sigma_i^2 \overset{ind}{\sim} N(\mu_i, \sigma_i^2),
\end{equation}
where $\mu_i$ follows a mixture model with a point mass at zero and $\sigma_i$ is drawn from an unspecified prior:
\beq\label{g.eqn}
\mu_i \overset{iid}{\sim} (1-\pi)\delta_0(\cdot)+\pi g_\mu(\cdot),\quad \sigma_i^2 \overset{iid}\sim g_\sigma(\cdot).
\eeq
In \eqref{g.eqn}, $\delta_0(\cdot)$ is a Dirac delta function indicating a point mass at 0 under the null hypothesis, while $g_{\mu}(\cdot)$ signifies that $\mu_i$ under the alternative is drawn from an unspecified distribution which is allowed to vary across $i$. In this work, we focus on a model where $\mu_i$ and $\sigma_i$ are not linked by a specific function. The more challenging situation where $\sigma_i$ may be informative for predicting $\mu_i$ is briefly discussed in Section \ref{open-issues:sec}.}

Following tradition in dealing with heteroscedasticity problems (e.g. \citealp{Xieetal12, Weietal18}), we assume that $\sigma_i$ are known. This simplifies the discussion and enables us to focus on key ideas. For practical applications, we use a consistent estimator of $\sigma_i$. The goal is to simultaneously test $m$ hypotheses:
\beq\label{hypos}
H_{0, i}: \mu_i=0 \quad \mbox{vs.} \quad H_{1, i}: \mu_i\neq 0; \quad i=1, \ldots, m.
\eeq
The multiple testing problem \eqref{hypos} is concerned with the simultaneous inference of $\pmb\theta=\{\theta_i=\mathbb I(\mu_i\neq 0): i=1, \ldots, m\}$, where $\mathbb I(\cdot)$ is an indicator function. The decision rule is represented by a binary vector $\bdelta = (\delta_i:  1\leq i\leq m)\in \{0, 1\}^m$, where $\delta_i = 1$ means that we reject $H_{0, i}$, and $\delta_i = 0$ means we do not reject $H_{0, i}$.
The false discovery rate (FDR) \citep{BenHoc95}, defined as
\beq \label{HATS:FDR}
\mbox{FDR}=E\left[\frac{\sum_i (1-\theta_i)\delta_i}{\max\{\sum_i \delta_i, 1\}}\right],
\eeq
is a widely used error criterion in large-scale testing problems. A closely related criterion is the marginal false discovery rate
\beq \label{mFDR}
\mbox{mFDR}=\frac{E\left\{\sum_i (1-\theta_i)\delta_i\right\}}{E\left(\sum_i \delta_i\right)}.
\eeq
The mFDR is asymptotically equivalent to the FDR for a general set of decision rules satisfying certain first- and second-order conditions on the number of rejections \citep{Basetal18}, including $p$--value based tests for independent hypotheses \citep{GenWas02} and weakly dependent hypotheses \citep{Stoetal04}. We shall show that our proposed data-driven procedure controls both the FDR and mFDR asymptotically; the main consideration of using the mFDR criterion is to derive optimality theory and facilitate methodological developments.

We use the expected number of true positives $\mbox{ETP}= E\left(\sum_{i=1}^m \theta_i\delta_i\right)$ to evaluate the power of an FDR procedure. Other power measures include the missed discovery rate (MDR, \citealp{Tayetal05}), average power (\citealp{BenHoc95, Efr07b}) and false negative rate or false non-discovery rate (FNR, \citealp{GenWas02, Sar02}). \cite{Caoetal13} showed that under the monotone likelihood ratio condition (MLRC), maximizing the ETP is equivalent to minimizing the MDR and FNR. The ETP is used in this article because it is intuitive and simplifies the theory. We call a multiple testing procedure \emph{valid} if it controls the FDR at the nominal level and \emph{optimal} if it has the largest ETP among all valid FDR procedures.

The building blocks for conventional multiple testing procedures are standardized statistics such as $Z_i$ or $P_i$. Let $\mu_i^*=\mu_i/\sigma_i$. The tacit rationale in conventional practice is that the simultaneous inference problem
\beq\label{hypos2}
H_{0,i}: \mu_i^*=0 \quad \mbox{vs.} \quad H_{1,i}: \mu_i^*\neq 0; \quad i=1, \ldots, m,
\eeq
is equivalent to the formulation \eqref{hypos}; hence the standardization step has no impact on multiple testing. However, this seemingly plausible argument, which only takes into account the null distribution, fails to consider the change in the structure of the alternative distribution. Next we present an example to illustrate the information loss and power distortion from standardizing.


\subsection{Data processing and power loss: an illustrative example}\label{example:sec}


The following diagram describes a data processing approach that is often adopted when performing hypothesis tests:
\beq\label{data-processing}
(X_i, \sigma_i) \quad \longrightarrow \quad Z_i=\frac{X_i}{\sigma_i} \quad \longrightarrow \quad P_i=2\Phi(-|Z_i|).
\eeq
We start with the full data consisting of $X_i$ and $\sigma^2_i=Var(X_i|\mu_i)$. The data is then standardized, $Z_i=X_i/\sigma_i$, and finally converted to a two-sided $p$-value, $P_i$. Typically these $p$-values are ordered from smallest to largest, a threshold is chosen to control the FDR, and 
hypotheses with $p$-values below the threshold are rejected.

Here we present a simple example to illustrate the information loss that can occur at each of these data compression steps. Consider a hypothesis testing setting with $H_{0,i}:\theta_i=0$ and the data coming from a normal mixture model, {where}
{\begin{equation}\label{toy.model}
\mu_i \overset{iid}{\sim} (1-\pi)\delta_0 + \pi\delta_{\mu_a}, \quad \sigma_i \overset{iid}{\sim} U[0.5,4].
\end{equation}}
\noindent {This is a special case of \eqref{g.eqn}, where $\mu_i$ are specifically drawn from a mixture of two point masses, and where we set  $\mu_a = 2.$}


We examine three possible approaches to controlling the FDR at $\alpha=0.1$. In the $p$-value approach we reject for all $p$-values below a given threshold. Note that, when the FDR is exhausted, this is the uniformly most powerful $p$-value based method \citep{GenWas02}, so is superior to, for example, the BH procedure. Alternatively, in the $z$-value approach we reject for all suitably small $\mathbb P(H_0|Z_i)$, which is in turn the most powerful $z$-value based method \citep{SunCai07}. Finally, in the full data approach we reject when $\mathbb P(H_0|X_i, \sigma_i)$ is below a certain threshold, which we show later is optimal given $X_i$ and $\sigma_i$. In computing the thresholds, we assume that there is an oracle knowing the alternative distribution; the formulas for our theoretical calculations are provided in Section \ref{formulas-example:sec} of the Appendix. For the model given by \eqref{toy.model} these rules correspond to:
\begin{eqnarray*}
	\pmb{\delta}^p & = & \{ \mathbb{I}(P_i\le 0.0006): 1\leq i\leq m \} =  \{ \mathbb{I}(|Z_i|\ge3.43): 1\leq i\leq m \}, \\
	\pmb{\delta}^z & = & \{ \mathbb{I}(\mathbb P(H_0|Z_i)\le 0.24): 1\leq i\leq m \}= \{ \mathbb{I}(Z_i\ge 3.13): 1\leq i\leq m \}, \\
	\pmb{\delta}^{\text{full}} & = & \{ \mathbb{I}(\mathbb P(H_0|X_i, \sigma_i)\le 0.28): 1\leq i\leq m \},
\end{eqnarray*}
with the thresholds chosen such that the FDR is exactly $10\%$ for all three approaches. However, while the FDRs of these three methods are identical, the average powers,
$\mbox{AP}(\pmb\delta)=\frac{1}{m\pi}\mathbb E\left(\sum_{i=1}^m \theta_i\delta_i\right)$, differ significantly:
\beq\label{power.com}
\mbox{AP}(\bdelta^p)=5.0\%, \quad \mbox{AP}(\bdelta^z)=7.2\%, \quad \mbox{AP}(\bdelta^{\text{full}})=10.5\%.
\eeq

\begin{figure}[t!]
	\vspace{-1cm}
	\begin{center}
		\includegraphics[width=0.95\textwidth]{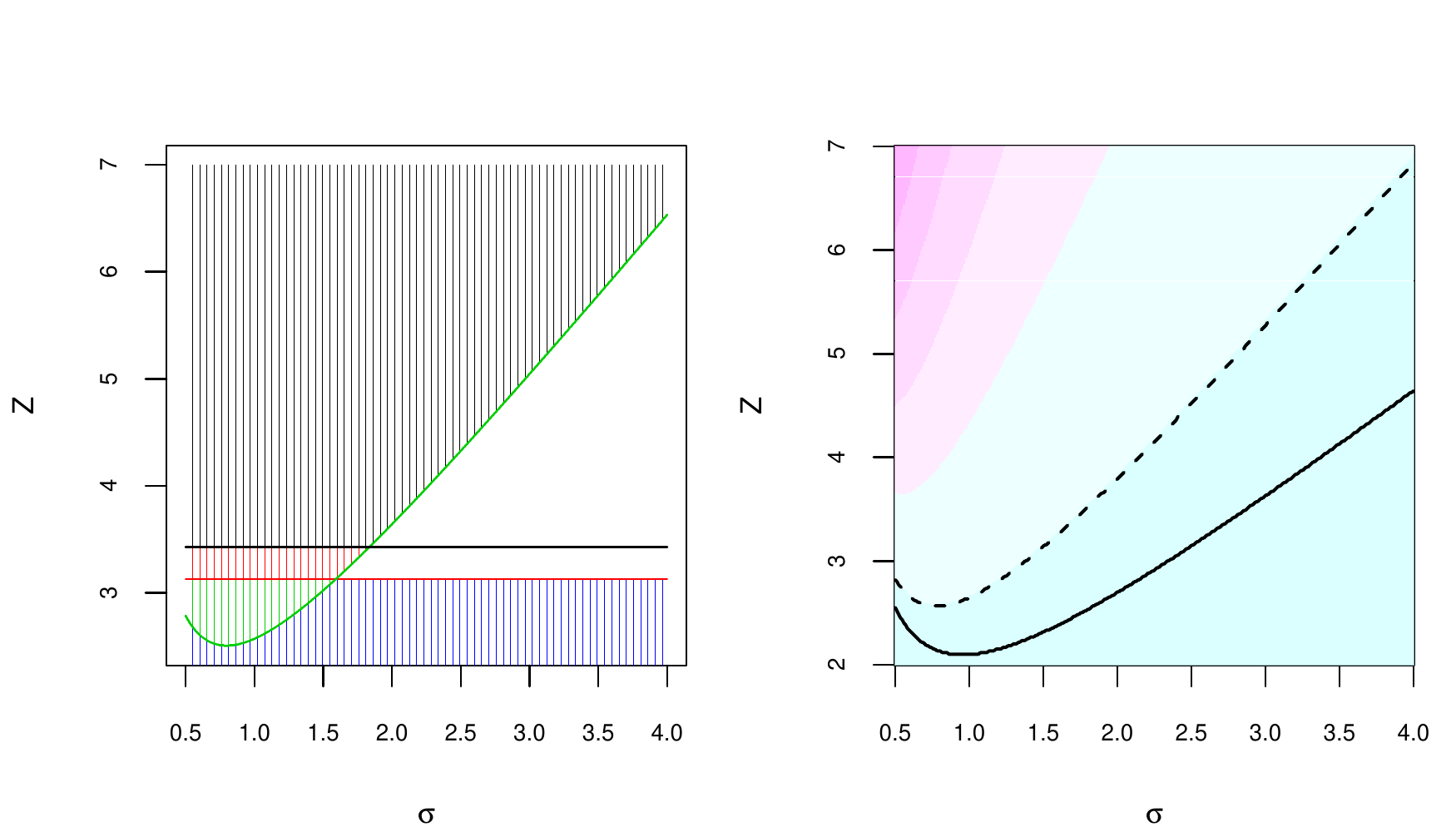}
		\caption{\small Left: Rejection regions for the $p$-value approach (black line), $z$-value approach (red line) and full data approach (green line) as a function of $Z$ and $\sigma$. Approaches reject for all points above their corresponding line. Right: Heat map of relative proportions (on log scale) of alternative vs null hypotheses for different $Z$ and $\sigma$. Blue corresponds to lower ratios and purple to higher ratios. The solid black line represents equal fractions of null and alternative, while the dashed line corresponds to three times as many alternative as null.}
		\label{toy.fig}
	\end{center}
\end{figure}

To better understand these differences consider the left hand plot in Figure~\ref{toy.fig}, which
illustrates the rejection regions for each approach as a function of $Z$ and $\sigma$\footnote{The $p$-value method will also reject for large negative values of $Z$ but, to keep the figure readable, we have not plotted that region.}. In the blue region all methods fail to reject the null hypothesis, while all methods reject in the black region. The green region corresponds to the space where the full data approach rejects the null while the other two methods do not. Alternatively, in the red region both the $z$-value and full data methods reject while the $p$-value approach fails to do so. Finally, in the white region the full data approach fails to reject while the $z$-value method does reject.

We first compare $\pmb\delta^z$ and $\pmb\delta^p$. Let $\pi^+$ and $\pi^-$ denote the proportions of positive effects and negative effects, respectively. Then $\pi^+=0.1$ and $\pi^-=0$. This asymmetry of the alternative distribution can be captured by $\pmb\delta^z$, which uses a one-sided rejection region. (Note that this asymmetric rejection region is not pre-specified but a consequence of theoretical derivation. In practice $\pmb\delta^z$ can be emulated by an adaptive $z$-value approach that is fully data-driven \citep{SunCai07}.) By contrast, $\pmb\delta^p$ enforces a two-sided  rejection region that is symmetrical about 0, trading off extra rejections in the region $Z_i\leq-3.43$ for fewer rejections in the region where $3.13\leq Z_i\leq 3.43$. As all nonzero effects are positive, negative $z$-values are highly unlikely to come from the alternative; this accounts for the 2.2\% loss in AP for the $p$-value method. Next consider $\pmb\delta^{\rm full}$ vs $\pmb\delta^z$. The full data approach trades off extra rejections in the green space for fewer rejections in the white space. This may seem like a sub-optimal trade-off given that the green space is smaller. However, the green space actually contains many more true alternative hypotheses. Approximately $3.8\%$ of the true alternatives occur in the green region as opposed to only $0.5\%$ in the white region, which accounts for the $3.3\%$ higher AP for the full data approach.

At first Figure~\ref{toy.fig} may appear counterintuitive. Why should we reject for low $z$-values in the green region but fail to reject for high $z$-values in the white region? The key observation here is that {\em not all $z$-values are created equal.} In the green region the observed data is far more consistent with the alternative hypothesis than the null hypothesis. For example, with $Z=4$ and $\sigma=0.5$ our observed $X$ is four standard deviations from the null mean but exactly equal to the alternative mean. Alternatively, while it is true that in the white region the high $z$-values suggest that the data are inconsistent with the null hypothesis, {\em they are also highly inconsistent with the alternative hypothesis.} For example, with $Z=4$ and $\sigma=2$ our observed $X$ is $8$, which is four standard deviations from the null mean, but also three standard deviations from the alternative mean. Given that $90\%$ of observations come from the null hypothesis, we do not have conclusive evidence as to whether this data is from the null or alternative. A $z$-value of $4$ with $\sigma=0.5$ is far more likely to come from the alternative hypothesis than is a $z$-value of $4$ with $\sigma=2$.

The right hand plot of Figure~\ref{toy.fig} makes this clear. Here we have plotted (on a log scale) the relative proportions of alternative vs null hypotheses for different $Z$ and $\sigma$. Blue corresponds to lower ratios and purple to higher ratios. The solid black line represents equal fractions of null and alternative, while the dashed line corresponds to three times as many alternative as null. Clearly, for the same $z$-value, alternative hypotheses are relatively more common for low $\sigma$ values. Notice how closely the shape of the dashed line maps the green rejection boundary in the left hand plot, which indicates that the full data method is correctly capturing the regions with most alternative hypotheses. By contrast, the $p$-value and $z$-value methods fail to correctly adjust for different values of $\sigma$.

\begin{figure}[t!]
	\vspace{0.1cm}
	\begin{center}
		\includegraphics[width=0.9\textwidth]{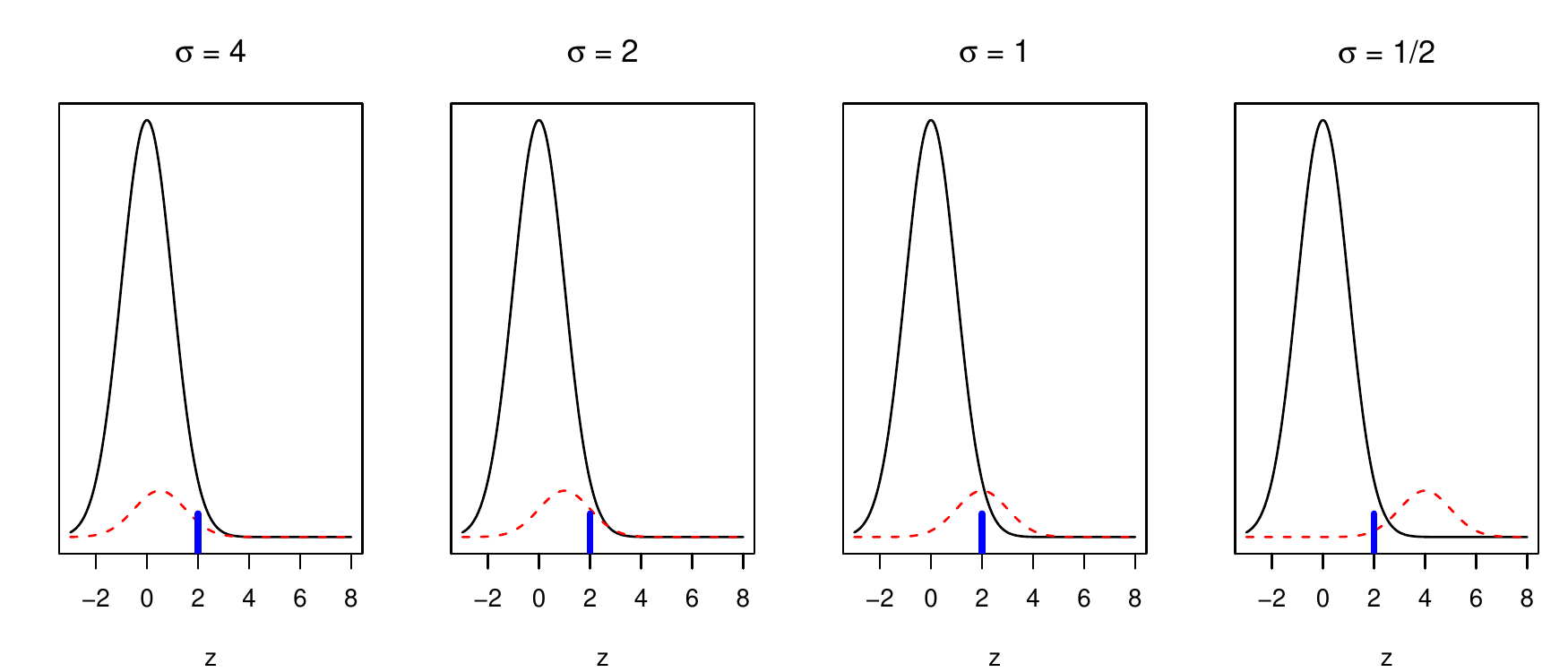}
		\caption{Plots of the density functions of $Z$ under the null hypothesis (black solid) and alternative hypothesis (red dashed) for different values of $\sigma$. The blue line represents an observation at $Z=2$.}
		\label{toy.fig2}
	\end{center}
\end{figure}

Figure~\ref{toy.fig2} provides one further way to understand the effect of standardizing the data. Here we have plotted the density functions of $Z$ under the null hypothesis (black solid) and alternative hypothesis (red dashed) for different values of $\sigma$. The densities have been multiplied by the relative probability of each hypothesis occurring so points where the densities cross correspond to an equal likelihood for either hypothesis. The blue line represents an observation, which is fixed at $Z=2$ in each plot. The alternative density is centered at $Z=2/\sigma$ so when $\sigma$ is large the standardized null and alternative are very similar, making it hard to know which distribution $Z=2$ belongs to. As $\sigma$ decreases the standardized alternative distribution moves away from the null and becomes more consistent with $Z=2$. However, eventually the alternative moves past $Z=2$ and it again becomes unclear which distribution our data belongs to. Standardizing means that the null hypothesis is consistent for all values of $\sigma$ but the alternative hypothesis can change dramatically as a function of the standard deviation.

To summarize, the information loss incurred in both steps of data processing \eqref{data-processing} reveals \emph{the essential role of the alternative distribution} in simultaneous testing. This structure of the alternative is not captured by the $p$-value, which is calculated only based on the null. Our result \eqref{power.com} in the toy example  shows that by exploiting (i) the overall asymmetry of the alternative via the $z$-value and (ii) the heterogeneity among individual alternatives via the full data, the average power of conventional $p$-value based methods can be doubled.

	\subsection{Heteroscadasticity and empirical null distribution}
	
	In the context of simultaneous testing with composite null hypotheses, \cite{SunMcL12} argued that the conventional testing framework, which involves rescaling or standardization, can become problematic:
	
	\medskip
	\small
	
	``In multiple testing problems where the null is simple
	($H_{0, i}: \mu_i=0$), the heteroscedasticity in errors can be removed
	by rescaling all $\sigma_i$ to 1. However, when the null is composite,
	such a rescaling step would distort the scientific question.''
	\medskip

	\normalsize
	
	\noindent \cite{SunMcL12} further proposed the concept of \emph{empirical composite null} as an extension of Efron's \emph{empirical null} (\citealp{Efr04}) for testing composite nulls $H_{0, i}: \mu_i\in [-a_0, a_0]$ under heteroscedastic models. It is important to note that the main message of this article, which focuses on the impact of heteroscedastiticy on the alternative instead of the null, is fundamentally different from that in \cite{SunMcL12}. In fact, we show that even when the null is simple, the heteroscedasticity still matters. Our finding, which somehow contradicts the above quotes, is more striking and even counter-intuitive. Moreover, we shall see that our data-driven HART procedure, which is based on Tweedie's formula (or the $f$-modeling approach, \citealp{Efr11}), is very different from the deconvoluting

\noindent kernel method (or $g$-modeling approach) in \cite{SunMcL12}\footnote{ The deconvoluting kernel method  has an extremely slow convergence rate. Our numerical studies show that the method in \cite{SunMcL12} only works for composite nulls where the uncertainties in estimation can be smoothed out over an interval $[-a_0, a_0]$. However, the deconvoluting method is highly unstable and does not work well when testing simple nulls $H_{0, i}: \mu_i=0$. Our numerical results show that the two-step method in Section \ref{sec:ddnew} works much better. }. The new two-step bivariate estimator in Section \ref{sec:ddnew} is novel and highly nontrivial; the techniques employed in the proofs of theory are also very different.

	\section{HART: Heteroscedasticity Adjusted Ranking and Thresholding}
\label{sec:method}

\setcounter{equation}{0}

The example in the previous section presents a setting where hypothesis tests based on the full data $(X_i,\sigma_i)$ can produce higher power than that from using the standardized data $Z_i$. In this section we formalize this idea and show that the result holds in general for heteroscedasticity problems. We first assume that the distributional information is known and derive an oracle rule based on full data in Section \ref{orac-rule:sec}. Section \ref{sec:ddnew} develops data-driven schemes and computational algorithms to implement the oracle rule. Finally theoretical properties of the proposed method are established in Section \ref{theory:sec}.


\subsection{The oracle rule under heteroscedasity}\label{orac-rule:sec}



Note that the models given by \eqref{x.eqn} and \eqref{g.eqn} imply that
\beq\label{x-rmix}
X_i |\sigma_i \overset{ind}{\sim} f_{\sigma_i}(x)=(1-\pi) f_{0,\sigma_i}(x)+\pi f_{1,\sigma_i}(x),
\eeq
where  $f_{0,\sigma}(x)=\frac{1}{\sigma}\phi(x/\sigma)$ is the null density, $f_{1, \sigma}(x)=\frac{1}{\sigma}\int \phi_{\sigma}\left(\frac{x-\mu}{\sigma}\right) g_\mu(\mu)d\mu$ is the alternative density, $\phi(x)$ is the density of a standard normal variable, and $f_{\sigma}(x)$ is the mixture density. For standardized data $Z_i=X_i/\sigma_i$, Model \ref{x-rmix} reduces to
\beq\label{z-rmix}
Z_i \overset{iid}{\sim} f(z)=(1-\pi) f_{0}(z)+\pi f_{1}(z),
\eeq	
where $f_0(z)=\phi(z)$, $f_1(z)$ is the non-null density, and $f(z)$ is the mixture density of the $z$-values.  As discussed previously, a standard approach involves converting the $z$-value to a two-sided $p$--value $P_i=2\Phi(-|Z_i|)$, where $\Phi(\cdot)$ is the standard normal cdf. The mixture model based on $p$-values is
\beq\label{p-rmix}
P_i \overset{iid}{\sim} g(p)=(1-\pi)\mathbb I_{[0,1]}(p)+\pi g_{1}(p), \quad \mbox{for } p\in[0,1],
\eeq
where $\mathbb I(\cdot)$ is an indicator function, and $g(\cdot)$ and $g_1(\cdot)$ are the mixture density and non-null density of the $p$-values, respectively. 	
Models \ref{z-rmix} and \ref{p-rmix} provide a powerful and flexible framework for large-scale inference and have been used in a range of related problems such as signal detection, sparsity estimation and multiple testing [e.g. \cite{Efretal01, Sto02, GenWas02, DonJin04, Newetal04, JinCai07}].

The oracle FDR procedure for Models~\ref{z-rmix} and \ref{p-rmix} are both known.
%
We first review the oracle $z$-value procedure \citep{SunCai07}. Define the local FDR (\citealp{Efretal01})
\beq\label{lfdr-st}
\mbox{Lfdr}_i=\mathbb P(H_0|z_i)=\mathbb P(\theta_i=0|z_i)=\frac{(1-\pi)f_0(z_i)}{f(z_i)}.
\eeq
Then \cite{SunCai07} showed that the optimal $z$-value FDR procedure is given by
\beq\label{or-z}
\bm\delta^z=[\mathbb I\{\mbox{Lfdr}(z_i)<c^*\}: 1\leq i\leq m],
\eeq
where $c^*$ is the largest Lfdr threshold such that $\mbox{mFDR}\leq \alpha$. Similarly, \cite{GenWas02} showed that the optimal $p$-value based FDR procedure is given by
\beq\label{or-p}
\bm\delta^p=[\mathbb I\{P_i<c^*\}: 1\leq i\leq m],
\eeq
where $c^*$ is the largest $p$-value threshold such that $\mbox{mFDR}\leq\alpha$.


Next we derive the oracle rule based on $m$ pairs $\{(x_i, \sigma_i): i=1, \ldots, m\}$. This new problem can be recast and solved in the framework of multiple testing with a covariate sequence. Consider Model \ref{x-rmix} and define the heterogeneity--adjusted significance index\footnote{Note that the oracle statistic $P(\theta_i=0|X_i, \sigma_i)$ is equivalent to $P(\theta_i=0|Z_i, \sigma_i)$ since the pairs $(X_i, \sigma_i)$ and $(Z_i, \sigma_i)$ contain the same amount of information. We use the pairs $(X_i, \sigma_i)$ in the next formula just to facilitate the estimation methodology.}
\beq\label{T-or}
T_i\equiv T(x_i, \sigma_i)= \mathbb P(\theta_i=0|x_i, \sigma_i) = \frac{(1-\pi)f_{0, \sigma_i}(x_i)}{f_{\sigma_i}(x_i)}.
\eeq
Let $Q(t)$ denote the mFDR level of the testing rule
$
[\mathbb I\{T_i<t\}: 1\leq i\leq m].
$
Then the oracle full data procedure is denoted
\beq\label{oracle-rule}
\bm\delta^{\text{full}}=[\mathbb I\{T_i<t^*\}: 1\leq i\leq m],
\eeq
where $t^*=\sup\{t: Q(t)\leq \alpha\}.$

The next theorem provides the key result showing that $\pmb\delta^{\text{full}}$ has highest power amongst all $\alpha$--level FDR rules based on $\{(x_i, \sigma_i): i=1, \cdots, m\}$.

\begin{theorem} \label{thm:optimal_or} Let $\mathcal D_\alpha$ be the collection of all testing rules based on $\{(x_i, \sigma_i): i=1, \ldots, m\}$ such that $\mbox{mFDR}_{\bm\delta}\leq \alpha$. Then $\mbox{ETP}_{\bm\delta}\leq \mbox{ETP}_{\bm\delta^{\text{full}}}$ for any $\bm\delta\in \mathcal D_\alpha$. In particular we have
	$$\mbox{ETP}_{\bm\delta^p}\leq \mbox{ETP}_{\bm\delta^z}\leq \mbox{ETP}_{\bm\delta^{\text{full}}}.$$
\end{theorem}

Based on Theorem~\ref{thm:optimal_or}, our proposed methodology employs a {\em heteroscedasticity--adjusted ranking and thresholding} (HART) rule that operates in two steps: first rank all hypotheses according to $T_i$ and then reject all hypotheses with $T_i\le t^*$. We discuss in Section~\ref{sec:ddnew} our finite sample approach for implementing HART using estimates for $T_i$ and $t^*$.


\subsection{Data-driven procedure and computational algorithms}\label{sec:ddnew}


We first discuss how to estimate $T_i$ and then turn to $t^*$. Inspecting $T_i$'s formula \eqref{T-or}, the null density $f_{0, \sigma_i}(x_i)$ is known and the non-null proportion $\pi$ can be estimated by $\hat \pi$ using existing methods such as Storey's estimator \citep{Sto02} or Jin-Cai's estimator \citep{JinCai07}. Hence we focus on the problem of estimating $f_{\sigma_i}(x_i)$.

There are two possible approaches for implementing this step. The first involves directly estimating $f_{\sigma_i}(x_i)$ while the second is implemented by first estimating $f_{1, \sigma_i}(x_i)$ and then computing the marginal distribution via
\beq\label{or:split-f}
\hat f_{\sigma_i}(x_i) = (1-\hat\pi)f_{0, \sigma_i}(x_i) + \hat\pi \hat f_{1, \sigma_i}(x_i).
\eeq
Our theoretical and empirical results strongly suggest that this latter approach provides superior results so we adopt this method.

\begin{remark}\rm{
		The main concern about the direct estimation of $f_{\sigma_i}(x_i)$ is that the tail areas of the mixture density are of the greatest interest in multiple testing but
		unfortunately the hardest parts to accurately estimate due to the few observations in the tails. The fact that $f_{\sigma_i}(x_i)$ appears in the denominator exacerbates the situation. The decomposition in \eqref{or:split-f} increases the stability of the density by incorporating a known part of null density. }
\end{remark}

Standard bivariate kernel methods \citep{Sil86, Wand94} are not suitable for estimating $f_{1, \sigma_i}(x_i)$ because, unlike a typical variable, $\sigma_i$ plays a special role in a density function and needs to be modeled carefully. \cite{FuJamesSun18} recently addressed a closely related problem using the following weighted bivariate kernel estimator:	
\beq\label{f-hat-2dim}
\hat{f}_{\sigma}^*(x)\coloneqq \sum_{j=1}^{m}\frac{\phi_{h_{\sigma}}(\sigma-\sigma_j)}{\sum_{j=1}^m\phi_{h_{\sigma}}(\sigma-\sigma_j)}\phi_{h_{xj}}(x-x_j),
\eeq
where ${\bm h}=(h_x, h_{\sigma})$ is a pair of bandwidths,  $\phi_{h_{\sigma}}(\sigma-\sigma_j)/\{\sum_{j=1}^m\phi_{h_{\sigma}}(\sigma-\sigma_j)\}$ determines the contribution of  $(x_j, \sigma_j)$ based on $\sigma_j$, $h_{xj}=h_x\sigma_j$ is a bandwidth that varies across $j$, and $\phi_{h}(z)= \frac{1}{\sqrt{2\pi}h}\exp\left\{-\frac{z^2}{2h^2}\right\}$ is a Gaussian kernel. The variable bandwidth $h_{xj}$ up-weights/down-weights observations corresponding to small/large $\sigma_j$; this suitably adjusts for the heteroscedasticity in the data.

Let $\mathcal M_1=\{i: \theta_i=1\}$. In the ideal setting where $\theta_j$ is observed one could extend \eqref{f-hat-2dim} to estimate $f_{1, \sigma_i}(x_i)$ via
\begin{equation}
\label{ideal.eq}
\tilde{f}_{1, \sigma}(x)= \sum_{j\in \mathcal M_1}\dfrac{\phi_{h_\sigma}(\sigma-\sigma_j)}{\sum_{k\in \mathcal M_1}\phi_{h_\sigma}(\sigma-\sigma_k)} \phi_{h_{xj}}(x-x_j).
\end{equation}
Given that $\theta_j$ is unknown, we cannot directly implement \eqref{ideal.eq}. Instead we apply a weighted version of \eqref{ideal.eq},
\begin{equation}
\label{real.f1}
\hat{f}_{1, \sigma_i}(x_i)=\sum_{j=1}^m   \dfrac{\hat{w}_j \phi_{h_\sigma}(\sigma_i-\sigma_j)}{\sum_{k=1}^{m} \hat{w}_k \phi_{h_\sigma}(\sigma_i-\sigma_k) }    \phi_{h_{xj}}(x_i-x_j)
\end{equation}
with weights $\hat w_j$ equal to an estimate of $P(\theta_j=1|x_j,\sigma_j)$. In particular we adopt a two step approach:
\begin{enumerate}
	\item Compute $\hat f_{1, \sigma_i}^{(0)}(x_i)$ via \eqref{real.f1} with initial weights $\hat{w}_j^{(0)} = (1-\hat T_j^{(0)})$ for all $j$, where $\hat T_j^{(0)}=\min\left\{\frac{(1-\hat \pi)f_{0, \sigma_j}(x_j)}{\hat{f}_{\sigma_j}^*(x_j)}, 1\right\}$, $\hat\pi$ is the estimated non-null proportion, and $\hat{f}_{\sigma_j}^*(x_j)$ is computed using \eqref{f-hat-2dim}.
	\item Compute $\hat f_{1, \sigma_i}^{(1)}(x_i)$ via \eqref{real.f1} with updated weights $\hat{w}_j^{(1)} = (1-\hat T_j^{(1)})$ where $$\hat T_j^{(1)}=\frac{(1-\hat\pi) f_{0,\sigma_j}(x_j)}{(1-\hat\pi)f_{0,\sigma_j}(x_j)+ \hat \pi \hat f_{1,\sigma_j}^{(0)}(x_j)}.$$
\end{enumerate}

This leads to our final estimate for $T_i=\mathbb P(H_0|x_i, \sigma_i)$:
$$\hat{T}_{i} =\hat T_i^{(2)}=\frac{(1-\hat\pi) f_{0,\sigma_i}(x_i)}{(1-\hat\pi)f_{0,\sigma_i}(x_i)+ \hat \pi \hat f_{1,\sigma_i}^{(1)}(x_i)}.$$
In the next section, we carry out a detailed theoretical analysis to show that both $\hat f_{\sigma_i}(x_i)$ and $\hat T_i$ are consistent estimators with $\mathbb E \| \hat{f}_{\sigma_i} - {f}_{\sigma_i}\|^2 = \mathbb E \int \{ \hat{f}_{\sigma_i}(x) - {f}_{\sigma_i}(x) \}^2 dx \rightarrow 0$ and  $\hat T_i\xrightarrow{P} T_i$, uniformly for all $i$.

To implement the oracle rule \eqref{oracle-rule}, we need to estimate the optimal threshold $t^*$, which can be found by carrying out the following simple stepwise procedure.
\begin{proc}[data-driven HART procedure]\label{proc:HATS.dd}
	Rank hypotheses by increasing order of $\hat{T}_{i}$. Denote the sorted ranking statistics $\hat{T}_{(1)} \leq \hdots \leq \hat{T}_{(m)}$ and $H_{(1)}, \hdots, H_{(m)}$ the corresponding hypotheses. Let
	$$
	k =\max \left\{j: \frac 1 j\sum_{i=1}^j \hat T_{(i)}\leq \alpha \right\}.
	$$
	Then reject the corresponding ordered hypotheses, $H_{(1)}, \hdots, H_{(k)}$.
\end{proc}

The idea of the above procedure is that if the first $j$ hypotheses are rejected, then the moving average $\frac 1 j\sum_{i=1}^j \hat T_{(i)}$ provides a good estimate of the false discovery proportion, which is required to fulfill the FDR constraint. Comparing with the oracle rule \eqref{oracle-rule}, Procedure \ref{proc:HATS.dd} can be viewed as its plug-in version:
\beq\label{proc:dd}
\pmb \delta^{dd}=\{\mathbb I(\hat T_i \leq \hat t^*): 1\leq i\leq m\}, \; \mbox{ where } \; \hat t^*=\hat T_{(k)}.
\eeq
The theoretical properties of Procedure 1 are studied in the next section.


\subsection{Theoretical properties of {Data-Driven} HART}\label{theory:sec}


In Section \ref{orac-rule:sec}, we have shown that the (full data) oracle rule  $\pmb\delta^{\rm full}$ \eqref{oracle-rule} is valid and optimal for FDR analysis. This section discusses the key theoretical result, Theorem 2, which  shows that the performance of  $\pmb\delta^{\rm full}$ can be achieved by its finite sample version $\pmb\delta^{\rm dd}$ \eqref{proc:dd} when $m\rightarrow \infty$. Inspecting \eqref{proc:dd}, the main steps involve showing that both $\hat T_i$ and $\hat t^*$ are ``close'' to their oracle counterparts. To ensure good performance of the proposed procedure, we require the following conditions.

\begin{description}	
	\item\textbf{(C1)} $supp(g_\sigma)\in (M_1,M_2)$ and $supp(g_\mu)\in (-M,M)$ for some $M_1>0$, $M_2<\infty$, $M<\infty$.
	
	\item\textbf{(C2)} {The kernel function $K$ is a positive, bounded and symmetric function satisfying $\int K(t)=1, \int tK(t)dt=0$ and $\int t^2 K(t)dt<\infty$. The density function $f_{\sigma}(t)$ has bounded and continuous second derivative and is square integrable. }
	\item\textbf{(C3)} The bandwidths satisfy $h_x=o\{(\log m)^{-1}\}$, $\lim\limits_{m\rightarrow\infty}mh_xh^2_\sigma=\infty$, $\lim\limits_{m\rightarrow\infty}m^{1-\delta}h_\sigma h^2_x=\infty$ and  $\lim\limits_{m\rightarrow\infty}m^{-\delta/2}h_\sigma^2h_x^{-1}\rightarrow 0$ for some $\delta>0$.
	\item \textbf{(C4)} $\hat{\pi} \overset{p}{\rightarrow} \pi$.
\end{description}	
\begin{remark}\rm{ For Condition (C2), the requirement on $f_\sigma$ is standard in density estimation theory, and the requirements on the kernel $K$ is satisfied by our choice of Gaussian kernel. Condition (C3) is satisfied by standard choices of bandwidths in \citet{Wand94} and \cite{Sil86}. Jin-Cai's estimator \cite{JinCai07} fulfills Condition (C4) in a wide class of mixture models. }
\end{remark}

Our theory is divided into two parts. The next proposition establishes the theoretical properties of the proposed density estimator $\hat f_{\sigma}$ and the plug-in statistic $\hat T_i$. The convergence of $\hat t^*$ to $t^*$ and the asymptotic properties of $\pmb\delta^{\rm dd}$ are established in Theorem \ref{thm:optimal_HATS}.

\begin{proposition}\label{t-hat:prop}
	Suppose Conditions (C1) to {(C4)} hold. Then $$\mathbb E \| \hat{f}_{\sigma} - {f}_{\sigma}\|^2 = \mathbb E \int \{ \hat{f}_{\sigma}(x) - {f}_{\sigma}(x) \}^2 dx \rightarrow 0,$$ where the expectation $\mathbb E$ is taken over $(\pmb X, \pmb \sigma, \pmb\mu)$. Further, we have  $\hat T_i\xrightarrow{P} T_i$.
\end{proposition}

Next we turn to the performance of our data-driven procedure $\pmb\delta^{\rm dd}$ when $m\rightarrow\infty$.  A key step in the theoretical development is to show that $\hat t^*\xrightarrow{P}t^*$, where $\hat t^*$ and $t^*$ are defined in \eqref{proc:dd} and \eqref{oracle-rule}, respectively.

\begin{theorem} \label{thm:optimal_HATS} Under the conditions in Proposition \ref{t-hat:prop}, we have $\hat t^*\xrightarrow{P}t^*$. Further, both the mFDR and FDR of $\pmb\delta^{\rm dd}$ are controlled at level $\alpha + o(1)$, and $ETP_{\bm \delta^{dd}}/ETP_{\bm \delta^{full}}=1+o(1)$.
\end{theorem}

In combination with Theorem \ref{thm:optimal_or}, these results demonstrate that the proposed  finite sample HART procedure (Procedure  \ref{proc:HATS.dd}) is asymptotically valid and optimal.


\section{Simulation}\label{sec:simu}


\setcounter{equation}{0}

We first describe the implementation of HART in Section \ref{sec:bw}. Section \ref{sec:simu.mixture} presents results for the general setting where $\sigma_i$ comes from a continuous density function. In Section \ref{simu-group:sec}, we further investigate the effect of heterogeneity under a mixture model where $\sigma_i$ takes on one of two distinct values. {Simulation results for additional settings, including a non-Guassian alternative, unknown $\sigma_i$, weak dependence structure, non-Gaussian noise, and estimated empirical null, are provided in Section E of the Supplementary Material.}


\subsection{Implementation of HART}\label{sec:bw}

The accurate estimation of $\hat T_i$ is crucial for ensuring good performance of the HART procedure. The key quantity is the bivariate kernel density estimator $\hat f_{1,\sigma}(x)$, which depends on the choice of tuning parameters $\bm{h}=(h_x, h_\sigma)$. Note that the ranking and selection process in Procedure \ref{proc:HATS.dd} only involves small $\hat T_i$. To improve accuracy, the bandwidth should be chosen based on the pairs $(x_i,\sigma_i)$ that are less likely to come from the null. We first implement Jin and Cai's method \citep{JinCai07} to estimate the overall proportion of non-nulls in the data, denoted $\hat{\pi}$.
We then compute $h_x$ and $h_\sigma$ by applying Silverman's rule of thumb \citep{Sil86} to the subset of the observations $\{x_i: P_i<\hat{\pi}\}$. {When implementing HART}, 
we first estimate $f_\sigma(x)$ using the data without $(X_i, {\sigma_i})$, and then plug-in the unused data $(X_i, {\sigma_i})$ to calculate $\hat T_i$. This 
method can increase the stability of the density estimator. As shown in the proof of Proposition 1, the asymptotic properties of $\hat T_i$ hold for both the regular and jacknifed approaches.

\subsection{Comparison in general settings}\label{sec:simu.mixture}

We consider simulation settings according to Models~\ref{x.eqn} and \ref{g.eqn}, where $\sigma_i$ are uniformly generated from $U[0, \sigma_{max}]$. We then generate $X_i$ from a two-component normal mixture model
$$X_i|\sigma_i\overset{iid}{\sim} (1-\pi)N(0, \sigma_i^2)+\pi N(2, \sigma_i^2).$$ In the first setting, we fix $\sigma_{max}=4$ and vary $\pi$ from 0.05 to 0.15. In the second setting, we fix $\pi=0.1$ and vary $\sigma_{max}$ from 3.5 to 4.5. {Five} methods are compared: the ideal full data oracle procedure (OR), the $z$-value oracle procedure of \citep{SunCai07} (ZOR), the Benjamini-Hochberg procedure (BH), {AdaPT \citep{LeiFit18} (AdaPT),} and the proposed data--driven HART procedure (DD). {Note that we do not include methods that explore the usefulness of sparsity structure \citep{Scottetal2015, BocaLeek2018, LiBar19, Caietal19} since the primary objective here is to incorporate structural information encoded in $\sigma_i$. Also, although  \citet{Ignetal16} mention the idea of using $\sigma_i$ as a covariate to construct weighted $p$-values, no guidance is given on how to do so, and since the way in which $\sigma_i$ are incorporated is particularly important (as illustrated by Section \ref{example:sec}), we exclude it.}

\begin{figure}[t!]
	\begin{center}
		\includegraphics[width=5.4in]{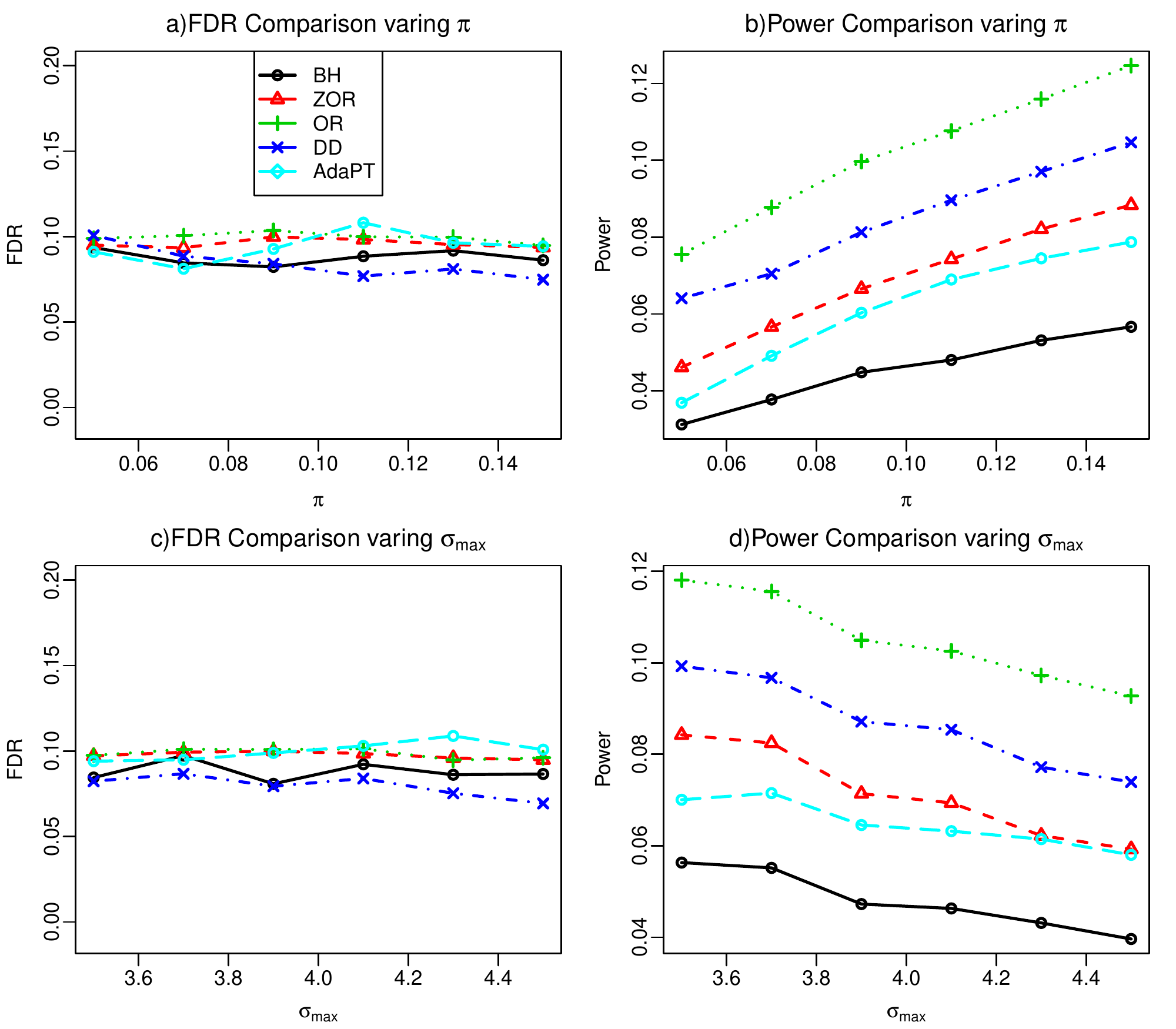}
		\caption{\label{plotcont} Comparison when $\sigma_i$ is generated from  a uniform distribution. We vary $\pi$ in the top row and  $\sigma_{max}$ in the bottom row. All methods control the FDR at the nominal level. DD has roughly the same FDR but higher power compared to ZOR in all settings. }
		\label{plot:mixture}
	\end{center}
\end{figure}

The nominal FDR level is set to $\alpha=0.1$. For each setting, the number of tests is $m = 20,000$. Each simulation is also run over $100$ repetitions. Then, the FDR is estimated as the average of the false discovery proportion $\mbox{FDP}(\bm{\delta}) = \sum_{i=1}^{m} \{(1-\theta_i) \delta_i\}/(\sum_{i=1}^{m}\delta_i \vee 1)$ and the average power is estimated as the average proportion of true positives that are correctly identified, $\sum_{i=1}^{m}(\theta_i\delta_i)/(mp)$, both over the number of repetitions. The results for differing values of $\pi$ and $\sigma_{max}$ are respectively displayed in the first and second rows of Figure~\ref{plotcont}.

Next we discuss some important patterns of the plots and provide interpretations. Panel (a) of Figure \ref{plotcont} shows that all methods appropriately control FDR at the nominal level, with DD being slightly conservative. Panel (b) illustrates the advantage of the proposed HART procedure over existing methods. When $\pi$ is small, the power of OR can be 60\% higher than ZOR. This shows that exploiting the structural information of the variance can be extremely beneficial. DD has lower power compared to OR due to the inaccuracy in estimation. However, DD still dominates ZOR and BH in all settings. We can also see that ZOR dominates BH and the efficiency gain increases as $\pi$ increases. To explain the power gain of ZOR over BH, let $\pi^+$ and $\pi^-$ denote the proportion of true positive signals and true negative signals, respectively. Then $\pi^+=\pi$ and $\pi^-=0$. This asymmetry can be captured by ZOR, which uses a one-sided rejection region. By contrast, BH adopts a two-sided symmetric rejection region. Under the setting being considered, the power loss due to the conservativeness of BH is essentially negligible, whereas the failure of capturing important structural information in the alternative accounts for most power loss.  From the second row of Figure \ref{plotcont}, we can again see that all methods control the FDR at the nominal level. OR dominates the other three methods in all settings. DD is less powerful than OR but has a clear advantage over ZOR with slightly lower FDR and higher power.
 {In most cases, AdaPT does perform better than BH. However, it is important to note that pre-ordering based on $\sigma_i$ is a suboptimal way for using side information. Moreover, the dominance of AdaPT over BH is not uniform, see section E.1 in the supplement for example. This shows that anti-informative pre-ordering based on $\sigma_i$ can lead to possible power loss for AdaPT. By contrast, HART  utilizes the side information in a principled and systematic way. It uniformly improves competitive methods.
Finally, it should be noted that incorporating side information comes with computational costs: conventional methods including BH and ZOR both run considerably faster than DD. However, DD runs faster than AdaPT.}

\subsection{Comparison under a two-group model}\label{simu-group:sec}

To illustrate the heteroscedasticity effect more clearly, we conduct a simulation using a simpler model where $\sigma_i$ takes on one of two distinct values. 
The example illustrates that the heterogeneity adjustment is more useful when there is greater variation in the standard deviations among the testing units.

\begin{figure}[t!]
	\begin{center}
		\includegraphics[width=5in]{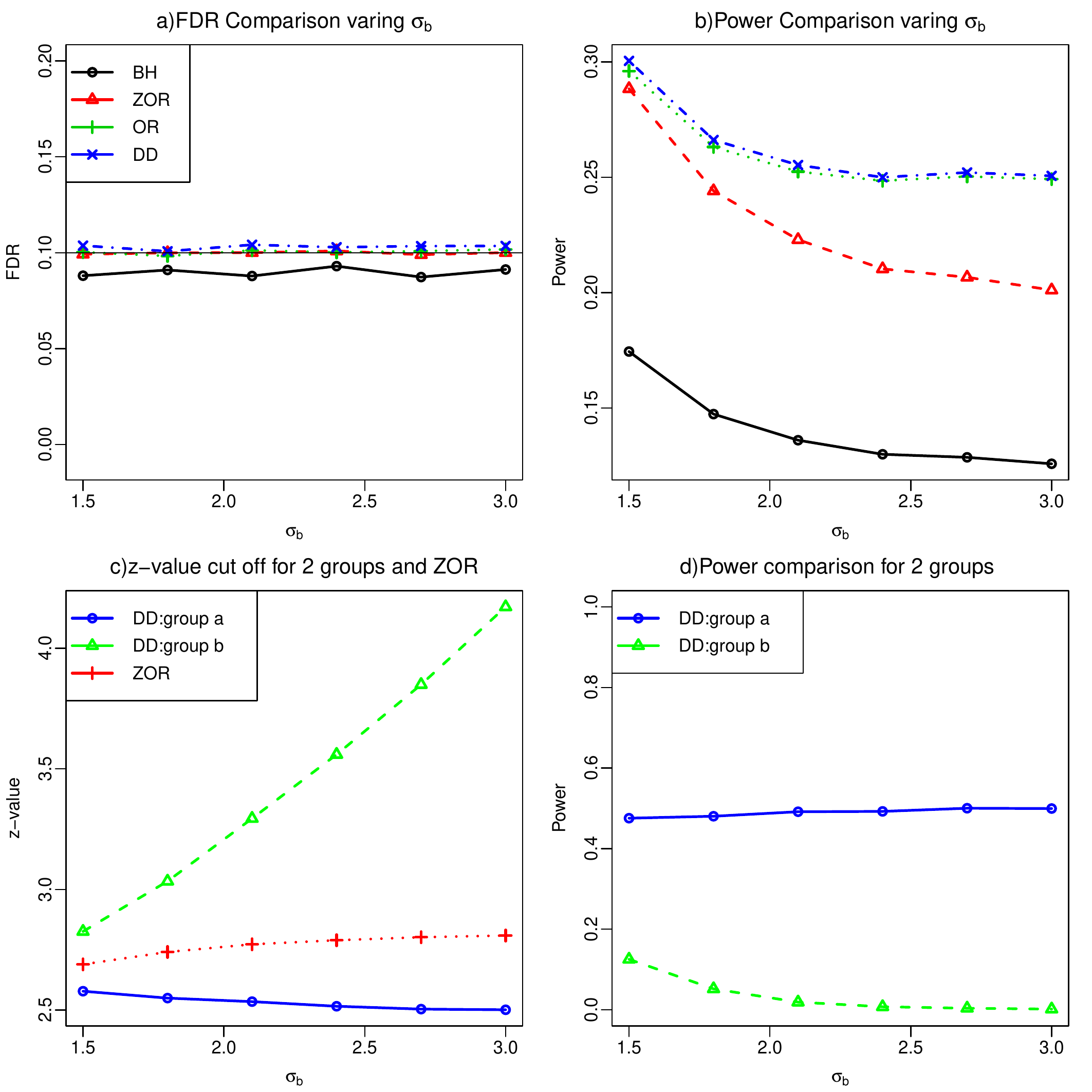}
		\caption {\label{plot2gp} Two groups with varying $\sigma_b$ from 1.5 to 3.  As $\sigma_b$ increases, the cut-off for group a decreases whereas the cut-off for group b increases. The power for tests in group b drops quickly as $\sigma_b$ increases.  This corroborates our calculations in the toy example in Section \ref{example:sec} and the patterns revealed by Figure \ref{toy.fig}.}
		\label{plot:2g}
	\end{center}
\end{figure}

Consider the setup in Models \ref{x.eqn} and \ref{g.eqn}. We first  draw $\sigma_i$ randomly from two possible values $\{\sigma_a, \sigma_b\}$ with equal probability, and then generate $X_i$ from a two-point normal mixture model
$X_i|\sigma_i\overset{iid}{\sim} (1-\pi)N(0, \sigma_i^2)+\pi N(\mu, \sigma_i^2)$.
In this simpler setting, it is easy to show that HART reduces to the CLfdr method in \cite{CaiSun09}, where the conditional Lfdr statistics are calculated for separate groups defined by $\sigma_a$ and $\sigma_b$. As previously, we apply BH, ZOR, OR and DD to data with $m=20,000$ tests and the experiment repeated on 100 data sets. We fix $\pi=0.1$, $\mu=2.5$, $\sigma_a=1$ and vary $\sigma_b$ from 1.5 to 3. The FDRs and powers of different methods are plotted as functions of $\sigma_b$, with results summarized in the first row of Figure \ref{plot:2g}. In the second row, we plot the group-wise $z$-value cutoffs and group-wise powers as functions of $\sigma_b$ for the DD method.

We can see that DD has almost identical performance to OR, and the power gain over ZOR becomes more pronounced as $\sigma_b$ increases. This is intuitive, because more variation in $\sigma$ tends to lead to more information loss in standardization. The bottom row shows the $z$-value cutoffs for ZOR and DD  for each group. We can see that in comparison to ZOR, which uses a single $z$-value cutoff, HART uses different cutoffs for each group. The $z$-value cutoff is bigger for the group with larger variance, and the gap between the two cutoffs increases as the degree of heterogeneity increases. In Panel d), we can see that the power of Group b decreases as $\sigma_b$ increases.  These interesting patterns corroborate those we observed in our toy example in Section \ref{example:sec}.

\section{Data Analysis}\label{sec:data}

This section compares the adaptive $z$-value procedure (AZ, the data-driven implementation of ZOR, \citet{SunCai07}), BH, and HART on a microarray data set. The data set measures expression levels of $12,625$ genes for patients with multiple myeloma, $36$ for whom magnetic resonance imaging (MRI) detected focal lesions of bone (lesions), and $137$ for whom  MRI scans could not detect focal lesions (without lesions) of bone \citep{Tianetal03}. For each gene, we calculate the differential gene expression levels ($X_i$) and standard errors ($S_i$). The FDR level is set at $\alpha = 0.1$.

\begin{figure}[t!]
	\begin{center}
		\includegraphics[width=5.5in]{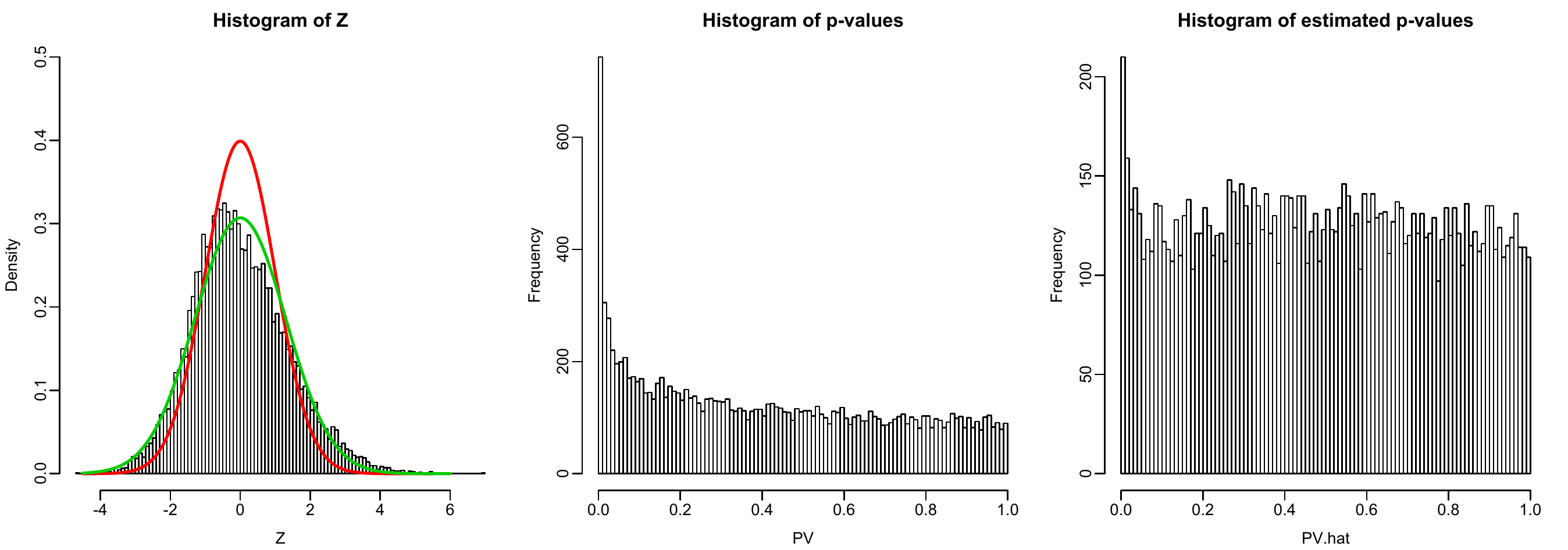}
		\caption{\label{plot:emnull} Left: histogram of $z$-values: the estimated empirical null $N(0, 1.3^2)$ (green line) seems to provide a better fit to the data compared to the theoretical null $N(0,1)$ (red line). Middle: histogram of original $p$-values. Right: histogram of estimated $p$-values based on the empirical null. The $z$-value histogram suggests that the theoretical null is inappropriate (too narrow, leading to too many rejections). The use of an empirical null corrects the non-uniformity of the histogram of the $p$-values.}
	\end{center}
\end{figure}

We first address two important practical issues. The first issue is that the theoretical null $N(0, 1)$ (red curve on the left panel of Figure \ref{plot:emnull}) is much narrower compared to the histogram of $z$-values. \cite{efron2004large} argued that a seemingly small deviation from the theoretical $z$-curve can lead to severely distorted FDR analysis. For this data set, the analysis based on the theoretical null would inappropriately reject too many hypotheses, resulting in a very high FDR. To address the distortion of the null, we adopted the \emph{empirical null} approach \citep{efron2004large} in our analysis. Specifically, we first used the middle part of the histogram, which contains 99\% of the data, to estimate the null distribution as $N(0, 1.30^2)$ [see \cite{efron2004large} for more details]. The new $p$-values are then converted from the $z$-values based on the estimated empirical null: $P_i^*=2\Phi^*(-2|Z_i|)$, where $\Phi^*$ is the CDF of a $N(0, 1.30^2)$ variable. We can see from Figure \ref{plot:emnull} that the empirical null (green curve) provides a better fit to the histogram of $z$-values. Another evidence for the suitability of the empirical null approach is that the histogram of the estimated $p$-values looks closer to uniform compared to that of original $p$-values. The uniformity assumption is crucial for ensuring the validity of $p$-value based procedures.

\begin{figure}[h!]
	\begin{center}
		\includegraphics[width=4.5in]{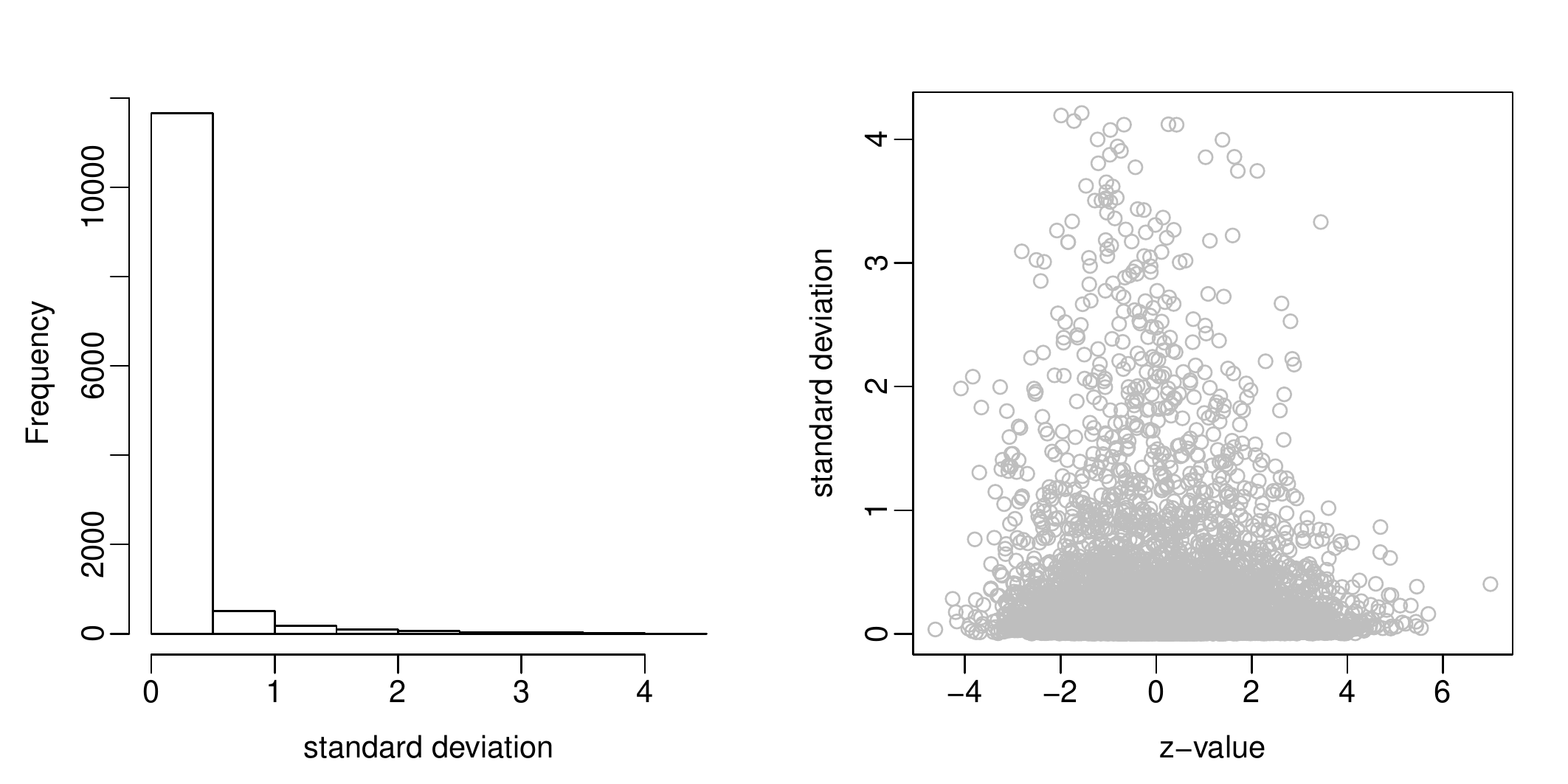}
		\caption{\label{plot:hist} Histogram of $S_i$ (left), scatter plot of $(Z_i, S_i)$ (right) }
	\end{center}
\end{figure}

\begin{figure}[!t]
	\begin{center}
		\includegraphics[width=5in]{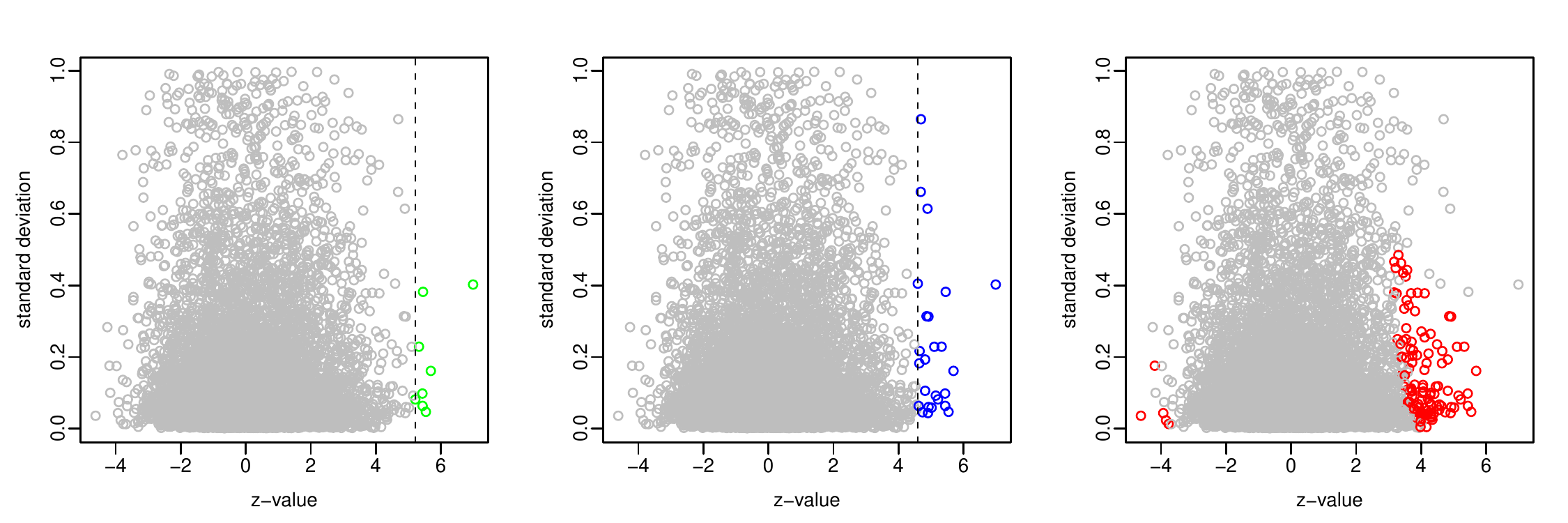}
		\caption{\label{plot:scat} Scatter plot of rejected hypotheses by each method. Green: BH, blue: AZ, red: HART. AZ and BH reject every hypothesis to the right of the dashed line. The rejection region for HART depends on both $z$ and $\sigma$.}
	\end{center}
\end{figure}

The second issue is the estimation of $f_\sigma(x)$, which usually requires a relatively large sample size to ensure good precision. Figure \ref{plot:hist} presents the histogram of $S_i$ and scatter plot of $S_i$ vs $Z_i$. Based on the histogram, we propose to only focus on data points with $S_i$ less than 1 (12172 out of 12625 genes are kept in the analysis) to ensure the estimation accuracy of $\hat T_i$. Compared to conventional approaches, there is no efficiency loss because no hypothesis with $S_i>1$ is rejected by BH at $\alpha=0.1$ -- note that the BH $p$-value cutoff is $6\times10^{-5}$, which corresponds to a $z$-value cutoff of 5.22; see also Figure \ref{plot:scat}. (If BH rejects hypotheses with large $S_i$, we recommend to carry out a group-wise FDR analysis, which first tests hypotheses at $\alpha$ in separate groups and then combines the testing results, as suggested by \cite{Efr08b}.)

Finally we apply BH, AZ and HART to the data points with $S_i<1$. BH uses the new $p$-values $P_i^*$ based on the estimated empirical null $N(0, 1.3^2)$. Similarly AZ uses Lfdr statistics where the null is taken as the density of a $N(0, 1.3^2)$ variable. When implementing HART, we estimate the non-null proportion $\pi$ using Jin-Cai's method with the empirical null taken as $N(0, 1.3^2)$. We further employ the jacknifed method to estimate $f_\sigma(x)$ by following the steps in Section \ref{sec:bw}. We summarize the number of rejections by each method in Table \ref{table:myeloma1} and display the testing results in Figure \ref{plot:scat}, where we have marked rejected hypotheses by each method using different colors.

HART rejects more hypotheses than BH and AZ. The numbers should be interpreted with caution as BH and AZ have employed the empirical null $N(0, 1.3^2)$ whereas HART has utilized  null density $N(0, \sigma_i^2)$ conditioned on individual $\sigma_i$ -- it remains an open issue how to extend the empirical null approach to the heteroscedastic case. Since we do not know the ground truth, it is difficult to assess the power gains. However, the key point of this analysis, and the focus of our paper, is to compare the \emph{shapes of rejection regions} to gain some insights on the differences between the methods. It can be seen that for this data set, the rejection rules of BH and AZ only depend on $Z_i$. By contrast, the rejection region for HART depends on both $Z_i$ and $S_i$. HART rejects more $z$-values when $S_i$ is small compared to BH and AZ. Moreover, HART does not reject any hypothesis when $S_i$ is large. This pattern is consistent with the intuitions we gleaned from the illustrative example (Figure \ref{toy.fig}) and the results we observed in simulation studies (Figure \ref{plot:2g}, Panel c).
\begin{table}
	\caption{\label{table:myeloma1} Numbers of genes (\% of total) that are selected by each method.}
	\centering
	\begin{tabular}{|l|l|l|l|}
		\hline
		$\alpha\text{-level} $& $\text{BH}$  & $ \text{AZ}$   & $ \text{HART}$\\
		\hline
		$0. 1$ & $8 ~ (0.07\%)$          &  $25~(0.2\%) $     &   $ 122 ~ (1\%)$   \\
		\hline
		
	\end{tabular}
	
\end{table}

\section{Discussion}\label{sec:extension}


\setcounter{equation}{0}

\subsection{Multiple testing with side information}

Multiple testing with side or auxiliary information is an important topic that has received much attention recently. The research directions are wide-ranging as there are various types of side information, which may be either extracted from the same data using carefully constructed auxiliary sequences or gleaned from secondary data sources such as prior studies, domain-specific knowledge, structural constraints and external covariates. The recent works by
\cite{Xiaetal19}, \cite{LiBar19} and \cite{Caietal19} have focused on utilizing side information that encodes the \emph{sparsity structure}. By contrast, our work investigates the impact of the \emph{alternative distribution}, showing that incorporating $\sigma_i$ can be extremely useful for improving the ranking and hence the power in multiple testing \footnote{
{A method is said to have better ranking if it discovers more true positives than its competitor at the same FDR level. Theorem 1 in Section 3.1 shows that the oracle HART procedure has the optimal ranking in the sense that it has the largest power among all FDR procedures at level $\alpha$}}.

In the context of FDR analysis, the key issue is that the hypotheses become unequal in light of side information. \cite{Efr08b} argued that ignoring the heterogeneity among study units may lead to FDR rules that are inefficient, noninterpretable and even invalid. We discuss two lines of work to further put our main contributions in context and to guide future research developments.

Grouping, pioneered by \cite{Efr08b}, provides an effective strategy for capturing the heterogeneity in the data. 
\cite{CaiSun09} showed that the power of FDR procedures can be much improved by utilizing new ranking statistics adjusted for grouping. Recent works along this direction, including \cite{Liuetal16}, \cite{BarRam17} and \cite{SarZha17}, develop general frameworks for dealing with a class of hierarchical and grouping structures. However, the groups can be characterized in many ways and the optimal grouping strategy still remains unknown. Moreover, discretizing a continuous covariate by grouping leads to loss of information. HART directly incorporates $\sigma_i$ into the ranking statistic and hence eliminates the need to define groups. 

Weighting is another widely used strategy for incorporating side information into FDR analyses \citep{BenHoc97, Genetal06, RoqVan09, Basetal18}. For example, when the sparsity structure is encoded by a covariate sequence, weighted $p$-values can be constructed to up-weight the tests at coordinates where signals appear to be more frequent \citep{Huetal12, Xiaetal19, LiBar19}.
However, the derivation of weighting functions for directly incorporating heteroscedasticity seems to be rather complicated \citep{Penetal11, Habetal17}. Notably, \cite{Hab17} developed novel weights for $p$-values as functions of a class of auxiliary parameters, including $\sigma_i$ as a special case, for a generic two-group mixture model. However, the formulation is complicated and the weights are hard to compute -- the methodology requires handling the derivative of the power function, estimating several unknown quantities and tuning a host of parameters.

\subsection{Open issues and future directions}\label{open-issues:sec}

We conclude the article by discussing several open issues. First, HART works better for large-scale problems where the  density with heteroscedastic errors can be well estimated. For problems with several hundred tests or less, $p$-value based algorithms such as BH or the WAMDF approach \citep{Hab17} are more suitable. The other promising direction for dealing with smaller-scale problems, suggested by \cite{CasRoq18}, is to employ spike and slab priors to produce more stable empirical Bayes estimates (with frequentist guarantees under certain conditions). Second, in practice the model given by \eqref{g.eqn} can be extended to
\beq\label{g2.eqn}
\mu_i|\sigma_i \overset{ind}{\sim} (1-\pi_{\sigma_i})\delta_0(\cdot)+\pi_{\sigma_i} g_\mu(\cdot|\sigma_i),\quad \sigma_i^2 \overset{iid}\sim g_\sigma(\cdot),
\eeq
where both the sparsity level and distribution of non-null effects depend on $\sigma_i$; this setting has been considered in a closely related work by \cite{Weietal18}. The heterogeneity-adjusted statistic is then given by
\beq\label{T-or2}
T_i=\mathbb P(\theta_i=0|x_i, \sigma_i) = \frac{(1-\pi_{\sigma_i})f_{0, \sigma_i}(x_i)}{f_{\sigma_i}(x_i)},
\eeq
where the varying proportion $\pi_{\sigma_i}$ indicates that $\sigma_i$ also captures the sparsity structure. This is possible, for example, in applications where observations from the alternative have larger variances compared to those from the null. An interesting, but challenging, direction for future research is to develop methodologies that can simultaneously incorporate both the sparsity and heterocedasticity structures into inference. Third, the HART-type methodology can only handle one covariate sequence $\{\sigma_i: 1\leq i\leq m\}$. It would be of great interest to develop new methodologies and principles for information pooling for multiple testing with several covariate sequences. Finally, our work has assumed that $\sigma_i$ are known in order to illustrate the key message (i.e. the impact of alternative distribution on the power of FDR analyses). Although this is a common practice, it is desirable to carefully investigate the impact of estimating $\sigma_i$ on the accuracy and stability of large-scale inference.
	
\bibliographystyle{apa}

\bibliography{myrefs}
\appendix

\newpage

\appendix

\begin{center}\LARGE
	Supplementary Material for ``Information Loss and Power Distortion from Standardizing in Multiple Hypothesis Testing''
\end{center}


\setcounter{equation}{0}

\section{Formulas for the Illustrative Example}\label{formulas-example:sec}

Consider Model	\ref{toy.model} in Section \ref{example:sec}. We derive the formulas for the oracle $p$-value, oracle $z$-value and oracle full data procedures.

\begin{itemize}
	\item $\pmb\delta^p$ corresponds to the thresholding rule $I(|Z_i|>t_p)$, where
	$$
	t_p=\inf\left\{t>0: \frac{2(1-\pi)\tilde{\Phi}(t)}{2(1-\pi)\tilde{\Phi}(t)+\pi \int \left\{ \tilde{\Phi}(t+\frac{\mu_a} \sigma)+\tilde{\Phi}(t-\frac{\mu_a} \sigma) \right\}d G(\sigma)} \leq\alpha\right\},
	$$
	with $\tilde{\Phi}$ being the survival function of the $N(0, 1)$ variable.
	
	\item $\pmb\delta^z$ is a one-sided thresholding rule of the form $I(Z_i>t_z)$, where
	$$
	t_z=\inf\left\{t>0: \frac{(1-\pi)\tilde{\Phi}(t)}{(1-\pi)\tilde{\Phi}(t)+\pi \int \tilde{\Phi}(t-\frac{\mu_a} \sigma) d G(\sigma)} \leq\alpha\right\}.
	$$
	
	\item $\pmb\delta^{\rm full}$ is of the form $I\{\mathbb P(\theta_i=0|x_i, \sigma_i)<\lambda\}$. It can be written as $I\{Z_i>t_{z, \sigma}(\lambda)\}$, where
	$$
	t_{z,\sigma}(\lambda)=\frac{\mu_a^2-2\sigma^2\log\left\{\frac{\lambda\pi}{(1-\lambda)(1-\pi)}\right\}}{2\mu_a\sigma}.
	$$
	Denote $\lambda^*$	the optimal threshold. Hence $\pmb\delta^{\rm full}$ is given by $I\{P(\theta_i=0|x_i,\sigma_i)<\lambda^*\}$, where
	$$
	\lambda^*=\sup\left[\lambda\in[0, 1]: \frac{(1-\pi)\int\tilde\Phi\{t_{z,\sigma}(\lambda)\}dG(\sigma)}{(1-\pi)\int\tilde\Phi\{t_{z,\sigma}(\lambda)\}dG(\sigma)+\pi\int \tilde\Phi\{t_{z,\sigma}(\lambda)-\frac {\mu_a}\sigma\}dG(\sigma)}  \right].
	$$
\end{itemize}

The optimal cutoffs can be solved numerically from the above. The powers are given by
\begin{eqnarray*}
	AP(\pmb\delta^p) & = & \int \left\{\tilde{\Phi}\left(t_p+\frac{\mu_a} \sigma\right)+\tilde{\Phi}\left(t_p-\frac{\mu_a} \sigma\right) \right\} d G(\sigma),
	\\
	AP(\pmb\delta^z) & = &  \int \tilde{\Phi}\left(t-\frac{\mu_a} \sigma\right) d G(\sigma), \\
	AP(\pmb\delta^{\rm full}) & = & \int \tilde\Phi\left\{t_{z,\sigma}(\lambda)-\frac {\mu_a}\sigma\right\}dG(\sigma).
\end{eqnarray*}


\section{Proofs of Theorems} \label{proofs-thms:sec}


\subsection{Proof of Theorem~\ref{thm:optimal_or}}\label{proof:thm.optimal_or}

We divide the proof into two parts. In Part (a), we establish two properties of the testing rule $\pmb\delta^{\rm full}(t)=\{\mathbb I (T_i<t): 1 \leq i \leq m\}$ for an arbitrary $0<t<1$. In Part (b) we show that the oracle rule $\pmb\delta^{\rm full}(t^*)$ attains the mFDR level exactly and is optimal amongst all FDR procedures at level $\alpha$.

\medskip

\noindent\textbf{Part (a).} Denote $\alpha(t)$ the mFDR level of $\pmb\delta^{\rm full}(t)$. We shall show that (i) $\alpha(t) < t$ for all $0<t<1$ and that (ii) $\alpha(t)$ is nondecreasing in $t$. Note that $\EE\left\{\sum_{i=1}^{m}(1-\theta_i)\delta_i\right\}= \EE_{\bm{X, \sigma}}\left(\sum_{i=1}^{m}T_i\delta_i\right)$. According to the definition of $\alpha(t)$, we have
\beq\label{eq:rewritemFdr}
\EE_{\bm{X, \sigma}}\left\{\sum_{i=1}^{m}\left\{T_i - \alpha(t)\right\}\mathbb{I}(T_i \leq t)\right\} = 0.
\eeq
We claim that $\alpha(t)< t$. Otherwise if $\alpha(t)\geq t$, then we must have  $T_i < t \leq \alpha(t)$. It follows that the LHS must be negative, contradicting \eqref{eq:rewritemFdr}.


Next we show (ii). Let $\alpha(t_j) = \alpha_j$. We claim that if $t_1 < t_2$, then we must have $\alpha_{1} \leq \alpha_{2}$. We argue by contradiction. Suppose that $t_1 < t_2$ but $\alpha_1 > \alpha_2$. Then 	
\begin{eqnarray*}
	(T_i - \alpha_2) \mathbb I(T_i < t_2) & = & (T_i - \alpha_1) \mathbb I(T_i < t_1)   +  (\alpha_1 - \alpha_2)  \mathbb I(T_i < t_1)+ (T_i - \alpha_2)  \mathbb I(t_1 \leq T_i < t_2) \\
	& \geq & (T_i - \alpha_1)  \mathbb I(T_i < t_1) + (\alpha_1 - \alpha_2)  \mathbb I(T_i < t_1)+ (T_i - \alpha_1)  \mathbb I(t_1 \leq T_i < t_2).
\end{eqnarray*}
It follows that $\mathbb E \left\{\sum_{i=1}^{m}(T_i - \alpha_2) \mathbb I(T_i < t_2) \right\} > 0$ since $\mathbb E \left\{\sum_{i=1}^{m}(T_i - \alpha_1) \mathbb I(T_i < t_1) \right\}=0$ according to \eqref{eq:rewritemFdr}, $\alpha_1 >\alpha_2$ and $T_i\geq t_1>\alpha_1$, contradicting \eqref{eq:rewritemFdr}. Hence we must have $\alpha_1 < \alpha_2.$

\medskip

\noindent\textbf{Part (b).} Let $\bar{\alpha} = \alpha(1)$. In Part (a), we show that $\alpha(t)$ is non--decreasing in $t$. It follows that for all $\alpha < \bar{\alpha}$, there exists a $t^*$ such that $t^*=\sup\{t: \alpha(t^*) = \alpha\}$. By definition, $t^*$ is the oracle threshold. Consider an arbitrary decision rule $\bm d= (d_{1}, \hdots, d_{m})\in \{0, 1\}^m$ such that $\mbox{mFDR}(\bm d) \leq \alpha$. We have $\mathbb {E} \left\{\sum_{i=1}^{m}(T_i - \alpha) \delta^{full}_i \right\} = 0 \text{ and } \mathbb E \left\{\sum_{i=1}^{m}(T_i - \alpha) d_i \right\} \leq 0.$ Hence
\beq\label{eq:or.power1}
\mathbb E \left\{\sum_{i=1}^{m}(\delta^{full}_i - d_i)(T_i - \alpha)  \right\} \geq 0.
\eeq
Consider transformation $f(x) = (x-\alpha)/(1-x)$. Note that $f(x)$ is monotone, we rewrite  $\delta^{full}_i = \mathbb I\left[\left\{ (T_i - \alpha)/(1-T_i)\right\} < \lambda\right]$, where $\lambda =(t^* - \alpha)/(1-t^*)$. In Part (a) we have shown that $\alpha < t_{OR} < 1$, which implies that $\lambda > 0$. Hence
\beq\label{eq:or.power2}
\mathbb E\left[\sum_{i=1}^{m}(\delta^{full}_i-d_i)\left\{(T_i-\alpha) - \lambda(1-T_i)\right\}\right] \leq 0.
\eeq
To see this, consider the terms where $\delta^{full}_i-d_i \neq 0$. Then we have two situations: (i) $\delta^{full}_i > d_i$ or (ii) $\delta^{full}_i < d_i$. In situation (i), $\delta^{full}_i = 1$, implying that $\left\{(T_i - \alpha)/(1-T_i)\right\} < \lambda$. In situation (ii), $\delta^{full}_i = 0$, implying that $\left\{(T_i - \alpha)/(1-T_i)\right\} \geq \lambda$. Therefore we always have $(\delta^{full}_i-d_i)\left\{(T_i-\alpha) - \lambda(1-T_i)\right\} \leq 0$. Summing over the $m$ terms and taking the expectation yield \eqref{eq:or.power2}. Combining \eqref{eq:or.power1} and \eqref{eq:or.power2}, we obtain
\[
0 \leq \mathbb E \left\{\sum_{i=1}^{m}(\delta^{full}_i - d_i)(T_i - \alpha)  \right\}
\leq \lambda \mathbb E \left\{\sum_{i=1}^{m}(\delta^{full}_i - d_i)(T_i - \alpha)  \right\}.
\]
Finally, since $\lambda > 0$, it follows that $\mathbb E \left\{\sum_{i=1}^{m}(\delta^{full}_i - d_i)(T_i - \alpha)  \right\} > 0$. Finally, we apply the definition of ETP to conclude that $\mbox{ETP}(\bm{\delta}^{full}) \geq \mbox{ETP}(\bm d)$ for all $\bm d \in \cal D_{\alpha}$.

\subsection{Proof of Theorem~\ref{thm:optimal_HATS}}\label{proof:thm.optimal_HATS}

We begin with a summary of notation used throughout the proof:

\begin{itemize}
	\item $Q(t) = m^{-1}\sum_{i=1}^m (T_i-\alpha) \mathbb{I}\{T_i < t\}$.
	\item $\hQ(t) = m^{-1}\sum_{i=1}^m (\hat T_i-\alpha) \mathbb{I}\{\hat T_i < t\}$.
	\item $Q_{\infty}(t) = E\{(\Tor-\alpha)\mathbb{I}\{\Tor<t\}\}$.
	\item $t_{\infty} = \sup\{t \in (0,1): Q_{\infty}(t) \leq 0\}$: the ``ideal'' threshold.
\end{itemize}

For $\Tor^{ (k)} < t< \Tor^{ (k+1)}$, define a continuous version of $\hQ(t)$ as
\[
\hQ_{C}(t) = \frac{t-\hTor^{ (k)}}{\hTor^{ (k+1)} - \hTor^{ (k)}} \hQ_{k} + \frac{\hTor^{ (k+1)}-t}{\hTor^{ (k+1)} - \hTor^{ (k)}} \hQ_{k+1},
\]
where $\hQ_{k} = \hQ\left(\hTor^{ (k)}\right)$. Since $\hQ_{C}(t)$ is continuous and monotone, its inverse $\hQ_{C}^{ -1}$ is well--defined, continuous and monotone. Next we show the following two results in turn: (i) $\hQ(t) \overset{p}\rightarrow Q_{\infty}(t)$ and (ii) $\hQ_{C}^{-1}(0) \overset{p}\rightarrow t_{\infty}$.

To show (i), note that $Q(t) \overset{p}\rightarrow Q_{\infty}(t)$ by the WLLN, so that we only need to establish that  $\hQ(t)-Q(t) \overset{p}\rightarrow 0$.
We need the following lemma, which is proven in Section \ref{proofs-lems:sec}.

\begin{lemma}\label{lemma:2}
	Let $U_i = (T_i-\alpha) \mathbb{I}(T_i < t)$ and $\hU_i = (\hat T_i-\alpha)\mathbb{I}\{\hat T_i < t\}$. Then $\mathbb E\left(\hU_i - U_i \right)^2 = o(1)$.
\end{lemma}

By Lemma \ref{lemma:2} and Cauchy-Schwartz inequality,  $\mathbb E\left\{\left(\hU_i-U_i\right)\left(\hU_j-U_j\right)\right\} = o(1)$. Let $S_m = \sum_{i=1}^m \left(\hU_i - U_i\right)$.
It follows that
$$
Var\left( m^{-1} S_m  \right) \leq m^{-2}\sum_{i=1}^{m} \mathbb E\left\{ \left( \hU_i - U_i\right)^2 \right\} +O\left(\frac{1}{m^2}\sum_{i,j:i\neq j} \mathbb E\left\{\left(\hU_i-U_i\right)\left(\hU_j-U_j\right)\right\}\right)= o(1).
$$
By Proposition \ref{t-hat:prop}, $\mathbb E(m^{-1}S_m)\rightarrow 0$, applying Chebyshev's inequality, we obtain $m^{-1}S_m = \hQ(t) - Q(t) \overset{p}\rightarrow 0$. Hence (i) is proved. Notice that $Q_{\infty}(t)$ is continuous by construction, we also have $\hQ(t)\overset{p}\rightarrow \hQ_{C}(t)$.

Next we show (ii). Since  $\hQ_{C}(t)$ is continuous, for any $\varepsilon>0$, we can find $\eta>0$ such that
$\left|\hQ_{C}^{ -1}(0) - \hQ_{C}^{ -1}\left\{\hQ_{C}^{}\left(t_{\infty}^{}\right)\right\}\right| < \varepsilon$
if
$\left|\hQ_{C}^{}\left(t_{\infty}^{}\right) \right|< \eta$. It follows that
\[
P\left\{ \left|\hQ_{C}^{}\left(t_{\infty}^{}\right) \right|> \eta\right\} \geq P\left\{\left|\hQ_{C}^{ -1}(0) - \hQ_{C}^{ -1}\left\{\hQ_{C}^{}\left(t_{\infty}^{}\right)\right\}\right| > \varepsilon\right\}.
\]
Proposition \ref{t-hat:prop} and the WLLN imply that $\hQ_{C}^{}(t) \overset{p}\rightarrow Q_{\infty}^{}(t).$ Note that $Q_{\infty}^{}\left(t_{\infty}^{}\right) = 0$. Then $
P\left(\left|\hQ_{C}^{}\left(t_{\infty}^{}\right) \right|>\eta\right) \rightarrow 0.$
Hence we have $
\hQ_{C}^{-1}(0) \overset{p}\rightarrow \hQ_{C}^{ -1} \left\{\hQ_{C}\left(t_{\infty}\right)\right\} = t_{\infty}$, completing the proof of (ii).

To show FDR$(\pmb{\delta}^{dd}) =$ FDR$(\pmb{\delta}^{full}) + o(1) =\alpha + o(1)$, we only need to show mFDR$(\pmb{\delta}^{dd}) =$ mFDR$(\pmb{\delta}^{full}) + o(1)$. The result then follows from the asymptotic equivalence of FDR and mFDR, which was proven in \cite{Caietal19}.

Define the continuous version of $Q(t)$ as $Q_{C}(t)$ and the  corresponding threshold as $Q_{C}^{ -1}(0)$. Then by construction, we have
\[
\bm{\delta}^{dd} = \left[\mathbb{I} \left\{\hat T_i \leq \hQ_{C}^{ -1}(0)\right\}: 1 \leq i \leq m\right] \quad \text{and} \quad \bm{\delta}^{full} = \left[\mathbb{I} \left\{T_i\leq Q_{C}^{ -1}(0)\right\}: 1 \leq i \leq m\right].
\]
Following the previous arguments, we can show that $
Q_{C}^{ -1}(0) \overset{p}\rightarrow  t_{\infty}$. It follows that $\hQ_{C}^{ -1}(0) = Q_{C}^{ -1}(0) + o_{p}(1).$ By construction $\mbox{mFDR}(\bm{\delta}^{full})=\alpha$. The mFDR level of $\bm{\delta}^{dd}$ is
\[
\mbox{mFDR}(\bm{\delta}^{dd}) = \frac{P_{H_0}\left\{\hat T_i \leq \hQ_{C}^{ -1}(0)\right\}}{P\left\{\hat T_i \leq \hQ_{C}^{ -1}(0)\right\}}.
\]
From Proposition 2, $\hat T_i \overset{p}\rightarrow T_i$. Using the continuous mapping theorem, $\mbox{mFDR}\left(\pmb{\delta}^{dd}\right) = \mbox{mFDR}\left(\pmb{\delta}^{full}\right) + o(1)=\alpha+o(1)$. The desired result follows.

Finally, using the fact that  $\hat T_i \overset{p}\rightarrow T_i$ and $\hQ_{C}^{ -1}(0)   \overset{p}\rightarrow Q_{C}^{ -1}(0)$, we can similarly show that
$$
\mbox{ETP}(\pmb{\delta}^{dd})/\mbox{ETP}(\pmb{\delta}^{full}) =1 + o(1).
$$


\section{Proof of Proposition \ref{t-hat:prop}}\label{proof-props:sec}



\subsection*{Summary of notation}


The following notation will be used throughout the proofs:
\begin{itemize}
	\item $\hat{f}^*_{\sigma}(x)= \sum_{j=1}^{m}\bigg\{\dfrac{\phi_{h_\sigma}(\sigma-\sigma_j)}{\sum_{i=1}^{m}\phi_{h_\sigma}(\sigma-\sigma_i)} \bigg\}\phi_{h_x\sigma_j}(x-x_j )$.
	\item $\fems(x)= \sum_{j=1}^{m}\bigg\{\dfrac{\phi_{h_\sigma}(\sigma-\sigma_j)\mathbb{I}(\theta_j=1)}{\sumind} \bigg\}\phi_{h_x\sigma_j}(x-x_j )$.
	\item  $\fne(x)=\sum_{j=1}^{m}\bigg\{\dfrac{\phi_{h_\sigma}(\sigma-\sigma_j)P(\theta_j=1|x_j,\sigma_j)}{\sumprob} \bigg\}\phi_{h_x\sigma_j}(x-x_j ) $.
	\item $\fem(x)=\sum_{j=1}^{m}\bigg\{\dfrac{\phi_{h_\sigma}(\sigma-\sigma_j)\hat{P}(\theta_j=1|x_j,\sigma_j)}{\sum_{i=1}^{m}\phi_{h_\sigma}(\sigma-\sigma_i)\hat{P}(\theta_i=1|x_i,\sigma_i)} \bigg\}\phi_{h_x\sigma_j}(x-x_j )$.
	\item $\hat{f}_\sigma(x)=(1-\hat{\pi})f_{0,\sigma}(x)+\hat{\pi}\fem(x)$.
\end{itemize}

The basic idea is that a consistent one--step estimator constructed via $\hat{f}^*_{\sigma}(x)$ leads to a consistent two--step estimator via $\hat{f}_{\sigma}(x)$.
By Condition \textit{(C4)} and the triangle inequality, it is sufficient to show that
\beq\label{eq:goal}
\mathbb{E}\int \left\{\hat{f}_{1,\sigma}(x)-f_{1,\sigma}(x)\right\}^2dx \rightarrow 0.
\eeq
Let $u_j=\dfrac{\phi_{h_\sigma}(\sigma-\sigma_j)}{\sum_{i=1}^{m}\phi_{h_\sigma}(\sigma-\sigma_i)}$. A direct consequence of condition (\textit{C1}) is
$0<\dfrac{C_1}{m}\leq \mathbb{E}u_j\leq \dfrac{C_2}{m}<\infty$ for some positive constants $C_1$ and $C_2$.
Let $C'=\min(1,C_1)$.Consider event
\beq\label{eventA}
\mathcal{A}=\left\{\left|\sum_{i=1}^{m} \theta_j-m\pi\right|<\frac{C'}{2}m\pi\right\}.
\eeq
By Hoeffding's inequality and Condition (\textit{C2}),
$
P(\mathcal{A}^C)O(h_x^{-2})\leq\exp(-C'^2m/2)O(h_x^{-2})\rightarrow 0.
$
Therefore it suffices to prove \eqref{eq:goal} under $\mathcal{A}$. We establish the result in three steps:
\begin{enumerate}
	\item $	\mathbb{E}\int\{\fems(x)-f_{1,\sigma}(x)\}^2dx\rightarrow 0 .$
	\item $	\mathbb{E}\int\{\fne(x)-\fems(x)\}^2dx\rightarrow 0  .$
	\item  $	\mathbb{E}\int \{\fem(x)-\fne(x)\}^2dx \rightarrow 0$.
\end{enumerate}
The proposition then follows from the triangle inequality.


\subsection{Proof of Step (a)}\label{step1:sec}

Let $b^*_j=\dfrac{\phi_{h_\sigma}(\sigma-\sigma_j)\mathbb{I}(\theta_j=1)}{\sum_{i=1}^{m}\phi_{h_\sigma}(\sigma-\sigma_i)\mathbb{I}(\theta_i=1)}$.
It is easy to show that
\begin{align*}
\left\{\hat{f}_{1,\sigma}^*(x)-f_{1,\sigma}(x)\right\}^2=\sum_{j=1}^{m}\sum_{k=1}^{m}b^*_jb^*_k\left\{\phi_{h_x\sigma_k}(x-x_k)-f_{1,\sigma}(x)\right\}\left\{\phi_{h_x\sigma_j}(x-x_j)-f_{1,\sigma}(x)\right\}.
\end{align*}
Under condition (\textit{C1}) and event $\mathcal{A}$, we have $\mathbb{E}(b^*_jb^*_k)=O(m^{-2})$.
Using standard arguments in density estimation theory (e.g. \cite{Wand94} page 21), and the fact that $\mathbb{E}{\sum_{j=1}^{m}(b^*_j)^2}=O(m^{-1}h^{-1}_\sigma), $ we have
$
\mathbb{E}\int\{\hat{f}_{1,\sigma}^*(x)-f_{1,\sigma}(x)\}^2dx=O\left\{(mh_\sigma h_x)^{-1}+h_x^4\right\}.
$
Under condition \textit{(C2)} and \textit{(C3)} the RHS $\rightarrow$ 0, establishing Step (a).

\subsection{Proof of Step (b)}\label{step2:sec}


Let $b_j=\dfrac{\phi_{h_\sigma}(\sigma-\sigma_j)P(\theta_j=1|x_j,\sigma_j)}{\sumprob}$. Then
\begin{eqnarray}\label{decomp1}
\left\{\fne(x)-\fems(x)\right\}^2 
& = & \sum_{j=1}^{m}(b^*_j-b_j)^2\phi^2_{h_x\sigma_j}(x-x_j) \nonumber \\ & & +\sum_{(j,k):j\neq k} (b^*_j-b_j)(b^*_k-b_k)\phi_{h_x\sigma_j}(x-x_j)\phi_{h_x\sigma_k}(x-x_k).
\end{eqnarray}

We first bound $\mathbb{E}(b^*_j-b_j)^2$. Write $\mathbb{E}(b^*_j-b_j)^2=\{\mathbb{E}(b^*_j-b_j)\}^2+Var(b^*_j-b_j)$.
It is clear that $\mathbb{E}(b^*_j)$ and $\mathbb{E}(b_j)$ are both $O(m^{-1})$. Hence $\{\mathbb{E}(b^*_j-b_j)\}^2=O(m^{-2})$. Next consider $Var(b^*_j-b_j)=Var(b^*_j)+Var(b_j)-2Cov(b^*_j,b_j)$. We have by condition (C3)
$$
Var(b^*_j)=Var\bigg\{\dfrac{\mathbb{I}(\theta_j=1)\phi_{h_\sigma}(\sigma-\sigma_j)}{\sumind}\bigg\}	\leq \mathbb{E}{(b^*_j)^2}=O(m^{-2}h^{-1}_\sigma).
$$
Similarly $Var(b_j)=O(m^{-2}h^{-1}_\sigma).$ It follows that $Cov(b^*_j,b_j)=O(m^{-2}h^{-1}_\sigma).$ Therefore $Var(b^*_j-b_j)=O(m^{-2}h^{-1}_\sigma)$ and $\mathbb{E}(b^*_j-b_j)^2=O(m^{-2}h^{-1}_\sigma)$. Using the fact that $\int\phi_{h_x\sigma_j}^2(x-x_j)dx=O(h_x^{-1})$, we have
\beq\label{step21}
\int\mathbb{E} \sum_{j=1}^{m}(b^*_j-b_j)^2\phi_{h_x\sigma_j}^2(x-x_j)dx=O\{(mh_xh_\sigma)^{-1}\}\rightarrow 0.
\eeq

Next we bound $\mathbb{E}\{(b^*_j-b_j)(b^*_k-b_k)\}$ for $j\neq k$. Consider the decomposition
\beq\label{decomp2}
\mathbb{E}\{(b^*_j-b_j)(b^*_k-b_k)\}=\mathbb{E}(b^*_j-b_j)\mathbb{E}(b^*_k-b_k)+Cov(b^*_j-b_j,b^*_k-b_k).
\eeq
Our goal is to show that $\mathbb{E}\{(b^*_j-b_j)(b^*_k-b_k)\}=O(m^{-3}h^{-2}_\sigma)+O(m^{-4}h^{-4}_\sigma)$. It suffices to show
\beq\label{eqbj}
\Et\{(b^*_j-b_j)(b^*_k-b_k)\}=O(m^{-3}h^{-2}_\sigma)+O(m^{-4}h^{-4}_\sigma).
\eeq
Observe that $Var\left\{\frac{1}{mh^{-1}_\sigma}\sum_{i=1}^{m}\phi_{h_{\sigma}}(\sigma-\sigma_j)\mathbb{I}(\theta_i=1)|\pmb{\sigma,x}\right\} = O(m^{-1})$ and
$$
\Et\left\{\frac{1}{mh^{-1}_\sigma}\sum_{i=1}^{m}\phi_{h_{\sigma}}(\sigma-\sigma_j)\mathbb{I}(\theta_i=1)\right\} = \frac{1}{mh^{-1}_\sigma}\sum_{i=1}^{m}\phi_{h_{\sigma}}(\sigma-\sigma_j)P(\theta_i=1|\sigma_i,x_i).
$$
Applying Chebyshev's inequality,
$$
\frac{1}{mh^{-1}_\sigma}\sum_{i=1}^{m}\phi_{h_{\sigma}}(\sigma-\sigma_j)\mathbb{I}(\theta_i=1)-\frac{1}{mh^{-1}_\sigma}\sum_{i=1}^{m}\phi_{h_{\sigma}}(\sigma-\sigma_j)P(\theta_i=1|\sigma_i,x_i)\xrightarrow{p}0.
$$
It follows that for any $\epsilon>0$,
$$
P\left\{\left|\sum_{i=1}^{m}\phi_{h_{\sigma}}(\sigma-\sigma_j)\mathbb{I}(\theta_i=1)-\sum_{i=1}^{m}\phi_{h_{\sigma}}(\sigma-\sigma_j)P(\theta_i=1|\sigma_i,x_i) \right|<\epsilon mh^{-1}_\sigma    \right\}\rightarrow 1.
$$
Under $\mathcal{A}$ defined in \eqref{eventA}, we have $\sum_{i=1}^{m}\phi_{h_{\sigma}}(\sigma-\sigma_j)\mathbb{I}(\theta_i=1)>h^{-1}_\sigma C_3m$ for some $C_3$, and
\beq\label{eq1}
P\left\{\sum_{i=1}^{m}\phi_{h_{\sigma}}(\sigma-\sigma_j)P(\theta_i=1|\sigma_i,x_i) <h_\sigma^{-1} C_3m \right\}\rightarrow 0.
\eeq
The boundedness of $b^*_j$ and $b_j$ and \eqref{eq1} imply that we only need to prove \eqref{eqbj} on the event
$\mathcal A^*=\left\{(\pmb x, \pmb\sigma): \sum_{i=1}^{m}\phi_{h_{\sigma}}(\sigma-\sigma_j)P(\theta_i=1|\sigma_i,x_i) \geq h^{-1}_\sigma C_3m\right\}.$
We shall consider $\Et(b^*_j-b_j)$ and $Cov(b^*_j-b_j,b^*_k-b_k|\pmb{\sigma},\pmb{x})$ in turn. Write
\begin{align*}
\Et (b^*_j) &=\Et\bigg\{\dfrac{\mathbb{I}(\theta_j=1)\phi_{h_\sigma}(\sigma-\sigma_j)}{\sumind}\bigg\}\\ &=P(\theta_j=1|x_j,\sigma_j)\Et\bigg\{\dfrac{\phi_{h_\sigma}(\sigma-\sigma_j)}{\sumind}\bigg\}\\
&+Cov\bigg\{\theta_j,\dfrac{\phi_{h_\sigma}(\sigma-\sigma_j)}{\sumind}\bigg|\pmb{\sigma,x}\bigg\}.	
\end{align*}
Let $Y=\dfrac{\phi_{h_\sigma}(\sigma-\sigma_j)}{\sumind}$. We state three lemmas that are proven in Section \ref{proofs-lems:sec}.

\bel\label{covbound}
Under event $\mathcal{A}^*$, we have $\Et(Y)-\Et(Y|\theta_j=1)=O(m^{-2})$ and\  $\Et(Y)-\Et(Y|\theta_j=0)=O(m^{-2}h^{-1}_\sigma)$.
\eel

\bel \label{diffexpratio}
Under event $\mathcal{A}^*$, we have $$\Et\bigg\{\dfrac{\phi_{h_\sigma}(\sigma-\sigma_j)}{\sumind}\bigg\}=\dfrac{\phi_{h_\sigma}(\sigma-\sigma_j)}{\sumprob}+O(m^{-2}h^{-2}_\sigma).$$
\eel


\bel\label{bouncov}
Under event $\mathcal{A}^*$, we have $Cov(b^*_j-b_j,b^*_k-b_k|\pmb{\sigma},\pmb{x})=O(m^{-3}h^{-2}_\sigma)$.
\eel

According to Lemma \ref{covbound}, we have
\begin{eqnarray}\label{eqcov}
& & Cov\bigg\{\theta_j,\dfrac{\phi_{h_\sigma}(\sigma-\sigma_j)}{\sumind}\bigg|\pmb{\sigma,x}\bigg\} \\ \nonumber
& = &\int \{P(\theta_j=0|x_j,\sigma_j)P(\theta_j=1|x_j,\sigma_j)\}\{y-\Et(Y)\}f_{Y|\theta_j=1,\pmb{\sigma,x}}(y)dy\\ \nonumber
&&-\int\{1-P(\theta_j=1|x_j,\sigma_j)\}^2\{y-\Et(Y)\}f_{Y|\theta_j=0,\pmb{\sigma,x}}(y)dy=O(m^{-2}h^{-1}_\sigma).
\end{eqnarray}
Together with Lemma \ref{diffexpratio}, we have
\beq\label{expratio}
\Et\bigg\{\dfrac{\mathbb{I}(\theta_j=1)\phi_{h_\sigma}(\sigma-\sigma_j)}{\sumind}\bigg\}=\dfrac{P(\theta_j=1|x_j,\sigma_j)\phi_{h_\sigma}(\sigma-\sigma_j)}{\sumprob}+O(m^{-2}h^{-2}_\sigma).
\eeq
It follows that $\mathbb{E}(b^*_j-b_j)=O(m^{-2}h^{-2}_\sigma)$ and $\mathbb{E}(b^*_j-b_j)\mathbb{E}(b^*_k-b_k)=O(m^{-4}h^{-4}_\sigma)$. The decomposition \eqref{decomp2} and Lemma \ref{bouncov} together imply $\mathbb{E}\{(b^*_j-b_j)(b^*_k-b_k)\}=O(m^{-3}h^{-2}_\sigma)+O(m^{-4}h^{-4}_\sigma)$. It follows that
\beq \label{qqq}
\int\mathbb{E}\sum_{(j,k):j\neq k} (b^*_j-b_j)(b^*_k-b_k)\phi_{h_x\sigma_j}(x-x_j)\phi_{h_x\sigma_j}(x-x_k)dx=O\{(mh_xh_\sigma^2)^{-1}+O\left\{(mh_\sigma)^{-2}\right\}\rightarrow 0.
\eeq
Combing \eqref{decomp1}, \eqref{step21} and \eqref{qqq}, we conclude that $\mathbb{E}\int\{\fne(x)-\fem^*(x)\}^2dx\rightarrow 0$.


\subsection{Proof of Step (c)}


Let $q_j=P(\theta_j=1|\sigma_j,x_j)$, $\hat{q}_j=\hat{P}(\theta_j=1|\sigma_j,x_j)=\min\left\{\dfrac{(1-\hat{\pi})f_{0,\sigma_j}(x_j)}{\hat{f}^{*}_{\sigma_j}(x_j)},1\right\}$ and
$
\fem(x)=\sum_{j=1}^{m}\dfrac{\phi_{h_\sigma}(\sigma-\sigma_j)\hat{q}_j}{\sum_{i=1}^m \phi_{h_\sigma}(\sigma-\sigma_i)\hat{q}_i}\phi_{h_x\sigma_j}(x-x_j).
$

Write $\hat{q}_j=q_j+a_j$, then $|a_j|\leq 1$ and $a_j=o_P(1)$. We have
\begin{eqnarray*}
	\mathbb{E}	\int \left\{\fem(x)-\fne(x)\right\}^2dx & = & O\left\{h_x^{-1}m^2\mathbb{E} \left( \dfrac{\phi_{h_\sigma}(\sigma-\sigma_j)\hat{q}_j}{\sum_{i=1}^{m}\phi_{h_\sigma}(\sigma-\sigma_i)\hat{q}_i}-\dfrac{\phi_{h_\sigma}(\sigma-\sigma_j)q_j}{\sum_{i=1}^{m} \phi_{h_\sigma}(\sigma-\sigma_i)q_i}   \right)^2 \right\}
	\\ & = & O\left\{h_x^{-1}h^2_\sigma\mathbb{E}a_j^2 \right\}.
\end{eqnarray*}
Next we explain why the last equality holds. Let $c_i=\phi_{h_\sigma}(\sigma-\sigma_i)h_\sigma$. Then
\begin{eqnarray*}
	& & \mathbb{E} \left\{ \dfrac{\phi_{h_\sigma}(\sigma-\sigma_j)\hat{q}_j}{\sum_{i=1}^{m}\phi_{h_\sigma}(\sigma-\sigma_i)\hat{q}_i}-\dfrac{\phi_{h_\sigma}(\sigma-\sigma_j)q_j}{\sum_{i=1}^{m} \phi_{h_\sigma}(\sigma-\sigma_i)q_i}   \right\}^2
	\\
	& = & \mathbb{E}\left\{\dfrac{a_j\sum_{i=1}^{m}\phi_{h_\sigma}(\sigma-\sigma_i)q_i-q_j\sum_{i=1}^{m}\phi_{h_\sigma}(\sigma-\sigma_i)a_i }     {\{\sum_{i=1}^{m}\phi_{h_\sigma}(\sigma-\sigma_i)\hat{q}_i\}\{\sum_{i=1}^{m}\phi_{h_\sigma}(\sigma-\sigma_i)q_i\}}  \right\}^2
	\\
	& = & h_\sigma^2\dfrac{1}{m^4}O\left[\mathbb{E}\left\{ a_j\sum_{i=1}^{m}c_iq_i-q_j\sum_{i=1}^{m}c_ia_i        \right\}^2 \right] \\ &=& \dfrac{h_\sigma^2}{m^4}O\left[\mathbb{E}\left\{m^2a_j^2-2ma_j\sum_{i=1}^{m}a_i+\left(\sum_{i=1}^{m}a_i\right)^2\right\}     \right] = \dfrac{h_\sigma^2}{m^2}O\left\{\mathbb{E}\left(a_j^2 \right)\right\}.
\end{eqnarray*}
The last line holds by noting that
$
\mathbb{E}\left(a_ja_i\right)\leq \sqrt{\mathbb{E}(a_j^2)\mathbb{E}(a_i^2)}=O\{\mathbb{E}(a_j^2)\}.
$

The next step is to bound $\mathbb{E}(a_j^2)$. Note that
$a_j=O\left\{\dfrac{ f_{0,\sigma_j}(x_j)\{  \hat{f}^{*}_{\sigma_j}(x_j)-f_{\sigma_j}(x_j)\}}{f_{\sigma_j}(x_j)\hat{f}^{*}_{\sigma_j}(x_j)}\right\}$. By the construction of $\hat{q}_j$, we have $\hat{f}_{\sigma_j}(x_j)\geq (1-\hat{\pi})f_{0,\sigma_j}(x_j)$.
Hence
$$
a_j=O\left\{ 1-\dfrac{\hat{f}^*_{\sigma_j}(x_j)}{f_{\sigma_j}(x_j)}   \right\} \;\mbox{ and } \;
\mathbb{E}(a_j^2)=O\left[\mathbb{E}\left\{1- \dfrac{\hat{f}^{*}_{\sigma_j}(x_j)}{f_{\sigma_j}(x_j)}  \right  \}^2   \right]. $$
Let
$
\mathcal K_j= \left(-\sigma_j\sqrt{\delta}\sqrt{\log m}-M,\sigma_j\sqrt{\delta}\sqrt{\log m}+M\right).
$
By the Gaussian tail bound, $P\left\{x_j\not\in \mathcal K_j \right \}=O(m^{-\delta/2}).$ By the boundedness of $a_j^2$ and the fact that $h_x^{-1}h_{\sigma}^2m^{-\delta/2}\rightarrow 0$ (Condition (\textit{C3})), we only need to consider $ \mathbb{E}\left[1-2\dfrac{\hat{f}^*_{\sigma_j}(x_j)}{f_{\sigma_j}(x_j)} +\left\{\dfrac{\hat{f}^*_{\sigma_j}(x_j)}{f_{\sigma_j}(x_j)} \right\}^2\bigg|x_j\right]$ for $x_j\in \mathcal K_j$.

Let
$f_{\sigma}(x_j)=\int \phi_\sigma(x)\left\{(1-\pi)\delta_0(x_j-x)+\pi g_\mu(x_j-x)\right\}dx$. Define a jacknifed version of $\hat{f}_{\sigma_j}^{*,(j)}$ that is formed without the pair $(\sigma_j,x_j)$. It follows that
$$\mathbb{E}\{\hat{f}^{*,(j)}_{\sigma_j}(x_j)|x_j\}=\int\int \phi_{\sqrt{\sigma^2+h_x^2\sigma^2_j}}(x)\left\{(1-\pi)\delta_0(x_j-x)+\pi g_\mu(x_j-x)\right\}g_\sigma(\sigma_j)d\sigma_jdx.$$
By the intermediate value theorem and Condition \textit{(C1)},
$$
\mathbb{E}\{\hat{f}^{*,(j)}_{\sigma_j}(x_j)|x_j\}=\int \phi_{\sqrt{\sigma^2+h_x^2c}}(x)\left\{(1-\pi)\delta_0(x_j-x)+\pi g_\mu(x_j-x)\right\} dx
$$ for some constant $c$. Next consider the ratio
\begin{align*}
\dfrac{\mathbb{E}\{\hat{f}^{*,(j)}_{\sigma_j}(x_j)|x_j\}}{f_{\sigma_j}(x_j)} &=\dfrac{\int \phi_{\sqrt{\sigma^2+h_x^2c}}(x)\left\{(1-\pi)\delta_0(x_j-x)+\pi g_\mu(x_j-x)\right\}dx}{\int \phi_\sigma(x)\left\{(1-\pi)\delta_0(x_j-x)+\pi g_\mu(x_j-x)\right\}dx}.
\end{align*}
By Condition \textit{(C1)}, $supp(g_\mu)\in(-M,M)$ with $M<\infty$, we have
$$
\inf_{x_j-M<x<x_j+M} \dfrac{\phi_{\sqrt{\sigma^2+h_x^2c}}(x)}{\phi_\sigma(x)} \leq	\dfrac{\mathbb{E}\{\hat{f}^{*,(j)}_{\sigma_j}(x_j)|x_j\}}{f_{\sigma_j}(x_j)}\leq \sup_{x_j-M<x<x_j+M} \dfrac{\phi_{\sqrt{\sigma^2+h_x^2c}}(x)}{\phi_\sigma(x)}.
$$
Note that $x_j\in (-\sigma_j\sqrt{\delta}\sqrt{\log m}-M,\sigma_j\sqrt{\delta}\sqrt{\log m}+M)$. The above infimum and supremum are taken over $x\in \mathcal K=(-\sigma_j\sqrt{\delta}\sqrt{\log m}-2M,\sigma_j\sqrt{\delta}\sqrt{\log m}+2M)$. Using Taylor expansion
$$
\dfrac{\phi_{\sqrt{\sigma^2+h_x^2c}}(x)}{\phi_\sigma(x)}=\dfrac{\sigma}{\sqrt{\sigma^2+h_x^2c}}\left[1+\sum_{k=1}^{\infty} \dfrac{1}{k!}\left\{\dfrac{h_x^2cx^2}{2(\sigma^2+h_x^2c)} \right\}^k \right],
$$
we have $\inf_{x\in\mathcal K}\dfrac{\phi_{\sqrt{\sigma^2+h_x^2c}}(x)}{\phi_\sigma(x)}=\dfrac{\sigma}{\sqrt{\sigma^2+h_x^2c}}= 1+O(h_x^2).$
Similarly,
\begin{align*}
\sup_{x\in\mathcal K}\dfrac{\sigma}{\sqrt{\sigma^2+h_x^2c}}\left\{1+\sum_{k=1}^{\infty} \dfrac{1}{k!}\left(\dfrac{h_x^2cx^2}{2(\sigma^2+h_x^2c)} \right)^k \right\}
&=1+O(h_x^2)+O\left[\sup\sum_{k=1}^{\infty} \dfrac{1}{k!}\left\{\dfrac{h_x^2cx^2}{2(\sigma^2+h_x^2c)} \right\}^k \right].
\end{align*}
It follows that
\begin{eqnarray}
\dfrac{\mathbb{E}\{\hat{f}_{\sigma_j}^{*,(j)}(x_j)|x_j\}}{f_{\sigma_j}(x_j)} & = & 1+O(h_x^2)+O\left\{\sup\sum_{k=1}^{\infty} \dfrac{1}{k!}\left(\dfrac{h_x^2cx^2}{2(\sigma^2+h_x^2c)} \right)^k\right\}, \label{eqn11} \\ \label{eqn12}
1-2\dfrac{\mathbb{E}\{\hat{f}_{\sigma_j}^{*,(j)}(x_j)|x_j\}}{f_{\sigma_j}(x_j)} & = & -1+O(h_x^2)+O\left\{\sup\sum_{k=1}^{\infty} \dfrac{1}{k!}\left(\dfrac{h_x^2cx^2}{2(\sigma^2+h_x^2c)} \right)^k \right\}.
\end{eqnarray}
Next consider
$
\mathbb{E}\left\{\dfrac{\hat{f}_{\sigma_j}^{*,(j)}(x_j)}{f_{\sigma_j}(x_j)} \bigg|x_j\right\}^2=\left[\dfrac{\mathbb{E}\{\hat{f}_{\sigma_j}^{*,(j)}(x_j)|x_j\}}{f_{\sigma_j}(x_j)} \right]^2+Var\left\{\dfrac{\hat{f}_{\sigma_j}^{*,(j)}(x_j)}{f_{\sigma_j}(x_j)}\bigg|x_j\right\}.
$
By the same computation on page 21 of \cite{Wand94},
$$Var\left\{\dfrac{\hat{f}_{\sigma_j}^{*,(j)}(x_j)}{f_{\sigma_j}(x_j)}\bigg|x_j\right\}=O\left\{(mh_x)^{-1} f_{\sigma_j}(x_j)^{-1} \right\}+o\left\{(mh_x)^{-1}f_{\sigma_j}(x_j)^{-2} \right\} .$$
Since $x_j\in\mathcal{K}_j$,  $f_{\sigma_j}(x_j)\geq C_3m^{-\delta/2}$ for some constant $C_3$, together with Condition (\textit{C3}), we have $h_x^{-1}Var\left\{\dfrac{\hat{f}_{\sigma_j}^{*,(j)}(x_j)}{f_{\sigma_j}(x_j)}\bigg|x_j\right\}=o(1).$ It follows from \eqref{eqn11} and \eqref{eqn12} that
\begin{eqnarray}\label{eqy}
& & h_x^{-1}-2h_x^{-1}\mathbb{E}\left\{\dfrac{\hat{f}^{*,(j)}_{\sigma_j}(x_j)}{f_{\sigma_j}(x_j)} \bigg|x_j\right\}+h_x^{-1}\mathbb{E}\left\{\dfrac{\hat{f}_{\sigma_j}^{*,(j)}(x_j)}{f_{\sigma_j}(x_j)} \bigg|x_j\right\}^2 \nonumber
\\	& = & O\left\{h_x+\sup\sum_{k=1}^{\infty} \dfrac{1}{k!}\left(\dfrac{h_x^{2k-1}c^kx^{2k}}{2^k(\sigma^2+h_x^2c)^k} \right) \right\}+o(1).
\end{eqnarray}
By condition \textit{(C2)} and the range of $x$, the RHS goes to 0.

Let $S_j=\sum_{i=1}^{m}\phi_{h_{\sigma}}(\sigma_j-\sigma_i)$ and $S_j^-=\sum_{i\neq j}^{m}\phi_{h_{\sigma}}(\sigma_j-\sigma_i).$
Some algebra shows $\hat{f}_{\sigma_j}^*(x_j)=\dfrac{S_j^-}{S_j}\hat{f}_{\sigma_j}^{*,(j)}(x_j)+\dfrac{1}{2S_j\pi h_\sigma h_x \sigma_j}$. We use the fact that $f_{\sigma_j}(x_j)\geq C_3m^{-\delta/2}$ for some constant $C_3$ and Condition \textit{(C3)} to claim that on $\mathcal A^*$,
\beq\label{eqx}
h_x^{-1}\dfrac{\mathbb{E}\{\hat{f}_{\sigma_j}^{*}(x_j)|x_j\}}{f_{\sigma_j}(x_j)}=h_x^{-1}\dfrac{\mathbb{E}\{\hat{f}_{\sigma_j}^{*,(j)}(x_j)|x_j\}}{f_{\sigma_j}(x_j)}+o(1).
\eeq
Similar computation shows that
\beq\label{eqz}
h_x^{-1}\mathbb{E}\left\{\dfrac{\hat{f}_{\sigma_j}^{*}(x_j)}{f_{\sigma_j}(x_j)} \bigg|x_j\right\}^2=h_x^{-1}\mathbb{E}\left\{\dfrac{\hat{f}_{\sigma_j}^{*,(j)}(x_j)}{f_{\sigma_j}(x_j)} \bigg|x_j\right\}^2+o(1).
\eeq
(\ref{eqy}), (\ref{eqx}) and (\ref{eqz}) together implies
$h_x^{-1}h^{2}_\sigma\mathbb{E}\{a_j^2|x_j\}\rightarrow 0$.
Hence
$
\mathbb{E}	\int \left\{\fem(x)-\fne(x)\right\}^2dx\rightarrow 0
$
and Step (c) is established. $\hat{T}_i\overset{p}{\rightarrow}T_i$ then follows from Lemma A.1 and Lemma A.2 in \cite{SunCai07}.


\section{Proof of Lemmas}\label{proofs-lems:sec}


\subsection{Proof of lemma \ref{lemma:2}}\label{app:proofLemma2}

Using the definitions of $\hU_i$ and $U_i$, we can show that $\left(\hU_i - U_i\right)^2 = \left(\hat T_i - T_i\right)^2\mathbb{I} \left(\hat T_i \leq t, T_i \leq t \right) + \left(\hat T_i-\alpha\right)^2\mathbb{I} \left(\hat T_i \leq t, T_i> t \right)+ \left(T_i-\alpha\right)^2\mathbb{I} \left(\hat T_i > t, T_i \leq t \right).$
Denote the three sums on the RHS as $I$, $II$, and $III$ respectively. By Proposition 2, $\mathbb{E}(I) = o(1)$. Let $\varepsilon > 0$. Consider
\begin{align*}
P\left(\hat T_i \leq t, T_i> t \right) &\leq P\left(\hat T_i \leq t, T_i\in \left(t, t+ \varepsilon \right) \right)+ P\left(\hat T_i \leq t, T_i\geq  t+ \varepsilon  \right) \\
&\leq P\left\{T_i \in \left(t, t+ \varepsilon \right)\right\} + P\left(\left|T_i - T_i\right| > \varepsilon \right)
\end{align*}
The first term on the right hand is vanishingly small as $\varepsilon \rightarrow 0$ because $\hTor^{i}$ is a continuous random variable. The second term converges to $0$ by Proposition 2. we conclude that $II = o(1)$. In a similar fashion, we can show that $III = o(1)$, thus proving the lemma.

\subsection{Proof of lemma \ref{covbound}}

Note that $\Et(Y|\theta_j=0)\geq \Et Y \geq \Et(Y|\theta_j=1)$. It is sufficient to bound $\Et(Y|\theta_j=0)-\Et(Y|\theta_j=1)$. The lemma follows by noting that
\begin{eqnarray*}
	&& \Et(Y|\theta_j=0)-\Et(Y|\theta_j=1) \\ & = & \Et \left\{\dfrac{\phi_{h_\sigma}(\sigma-\sigma_j)}{\sum_{i\neq j}\phi_{h_\sigma}(\sigma-\sigma_i)\theta_i}\right\}-\Et \left\{\dfrac{\phi_{h_\sigma}(\sigma-\sigma_j)}{\sum_{i\neq j}\phi_{h_\sigma}(\sigma-\sigma_i)\theta_i+\phi_{h_\sigma}(\sigma-\sigma_j)}\right\}\\
	& = & \Et \left\{\dfrac{\phi^2_{h_\sigma}(\sigma-\sigma_j)}{\{\sum_{i\neq j}\phi_{h_\sigma}(\sigma-\sigma_i)\theta_i\}\{\sum_{i\neq j}\phi_{h_\sigma}(\sigma-\sigma_i)\theta_i+\phi_{h_\sigma}(\sigma-\sigma_j)\}}\right\}\\
	&\leq & \Et\left\{\dfrac{\phi^2_{h_\sigma}(\sigma-\sigma_j)}{(\sum_{i\neq j}\phi_{h_\sigma}(\sigma-\sigma_i)\theta_i)^2}\right\}=O(m^{-2}h^{-1}_\sigma).
\end{eqnarray*}

\subsection{Proof of lemma \ref{diffexpratio}}
Let $Z=\sumind$,
We expand $\dfrac{1}{Z}$ around $\Et(Z)$ and take expected value:
\begin{equation*}
\Et\bigg(\dfrac{1}{Z}\bigg)=\Et \left[ \dfrac{1}{\Et(Z)}-\dfrac{1}{\{\Et(Z)\}^2}(Z-\Et Z)+ \sum_{k=3}^{\infty}\dfrac{(-1)^{k-1}}{\{\Et(Z)\}^k}(Z-\Et Z)^{k-1}\right].
\end{equation*}
The series converges on $\mathcal{A}$. Moreover, using normal approximation of binomial distribution, it can be shown that $\Et(Z-\Et Z)^k=O((mh_\sigma^{-1})^{k/2})$. The lemma follows by noting that
$\Et\left(Z^{-1}\right)=\{\Et(Z)\}^{-1}+O(m^{-2}h_\sigma^2)$.

\subsection{Proof of lemma \ref{bouncov}}

Consider $b^*_j=\dfrac{\phi_{h_\sigma}(\sigma-\sigma_j)\mathbb{I}(\theta_j=1)}{\sum_{i=1}^{m}\phi_{h_\sigma}(\sigma-\sigma_i)\mathbb{I}(\theta_i=1)}$ defined in Section \ref{step1:sec}. Let $\tilde{b}_j=\dfrac{\theta_j}{\sum_{i=1}^{m}\theta_i}$. By Condition (\textit{C1}), $Cov(b^*_j,b^*_k|\pmb{\sigma},\pmb{x})=O\{h_\sigma^{-2}Cov(\tilde{b}_j,\tilde{b}_k|\pmb{\sigma},\pmb{x})\} .$
Note that
$
Cov(\tilde{b}_j,\tilde{b}_k|\pmb{\sigma},\pmb{x})=\Et(\tilde{b}_j\tilde{b}_k)-\Et(\tilde{b}_j)\Et(\tilde{b}_k).
$
Using similar argument as in the proof for  (\ref{expratio}), we have
$$\Et(\tilde{b}_j)=\dfrac{P(\theta_j=1|\pmb{\sigma},\pmb{x})}{\sum_{i=1}^{m}P(\theta_i=1|\pmb{\sigma},\pmb{x})}+O(m^{-2})\; \mbox{and} \;
\Et(\tilde{b}_k)=\dfrac{P(\theta_k=1|\pmb{\sigma},\pmb{x})}{\sum_{i=1}^{m}P(\theta_i=1|\pmb{\sigma},\pmb{x})}+O(m^{-2}).
$$
It follows that
$
\Et(\tilde{b}_j)\Et(\tilde{b}_k)=\left\{  \dfrac{P(\theta_j=1|\pmb{\sigma},\pmb{x})}{\sum_{i=1}^{m}P(\theta_i=1|\pmb{\sigma},\pmb{x})} \right\}\left\{\dfrac{P(\theta_k=1|\pmb{\sigma},\pmb{x})}{\sum_{i=1}^{m}P(\theta_i=1|\pmb{\sigma},\pmb{x})} \right\}+O(m^{-3}).
$
Next we compute $\Et(\tilde{b}_j\tilde{b}_k)$. Note that
$
\Et(\tilde{b}_j\tilde{b}_k)=P(\theta_j=1|\pmb{\sigma},\pmb{x})\Et\left\{ \dfrac{\theta_k}{(\sum_{i=1}^{m}\theta_i)^2}\bigg| \theta_j=1   \right\}.$
Using similar arguments as the proof for (\ref{eqcov}), we have
$Cov\left\{\theta_k, \dfrac{1}{(\sum_{i=1}^{m}\theta_i)^2}\bigg| \theta_j=1,\pmb{\sigma},\pmb{x} \right\}=O(m^{-3}).
$
Let $\pmb\theta_{-k}=(\theta_j: 1\leq j\leq m, j\neq k)$. We have
$$
\Et\left\{ \dfrac{\theta_k}{(\sum_{i=1}^{m}\theta_i)^2}\bigg| \theta_j=1   \right\}=P(\theta_k=1|\pmb{\sigma},\pmb{x})\mathbb{E}_{\pmb\theta_{-k}}\left\{ \dfrac{1}{(\sum_{i=1}^{m}\theta_i)^2}\bigg| \theta_j=1,\pmb{\sigma},\pmb{x}   \right\}+O(m^{-3}).
$$
Using similar arguments in Lemmas \ref{diffexpratio} and \ref{covbound}, we have
$$
\mathbb{E}_{\pmb\theta_{-k}}\left\{ \dfrac{1}{(\sum_{i=1}^{m}\theta_i)^2}\bigg| \theta_j=1,\pmb{\sigma},\pmb{x}   \right\}=\dfrac{1}{\Et(\sum_{i=1}^{m}\theta_i)^2}+O(m^{-3}).
$$
In the previous equation, the conditional expectation $\mathbb{E}_{\pmb\theta_{-k}}$ can be replaced by $\mathbb{E}_{\pmb\theta}$ because the term $\theta_k$ only affects the ratio by a term of order $O(m^{-3})$.
Note that $\Et(\sum_{i=1}^{m}\theta_i)^2=\left\{\Et(\sum_{i=1}^{m}\theta_i)\right\}^2+Var(\sum_{i=1}^{m}\theta_i|\pmb{\sigma},\pmb{x}) $ and $Var(\sum_{i=1}^{m}\theta_i|\pmb{\sigma},\pmb{x})\leq m$. We have
$$
\Et\left\{ \dfrac{\theta_k}{(\sum_{i=1}^{m}\theta_i)^2}\bigg| \theta_j=1   \right\}=\dfrac{P(\theta_k=1|\pmb{\sigma},\pmb{x})}{\left\{ \sum_{i=1}^{m}P(\theta_i=1|\pmb{\sigma},\pmb{x})\right\}^2} +O(m^{-3}).
$$
Finally, the lemma follows from the fact that
$$Cov(\tilde{b}_j,\tilde{b}_k|\pmb{\sigma},\pmb{x})=\Et(\tilde{b}_j\tilde{b}_k)-\Et(\tilde{b}_j)\Et(\tilde{b}_k)=O(m^{-3}).$$

\section{Supplementary Numerical Results}\label{extranum}

\subsection{Non-Gaussian alternative}
\begin{figure}[t!]
	\begin{center}
		\includegraphics[width=5in]{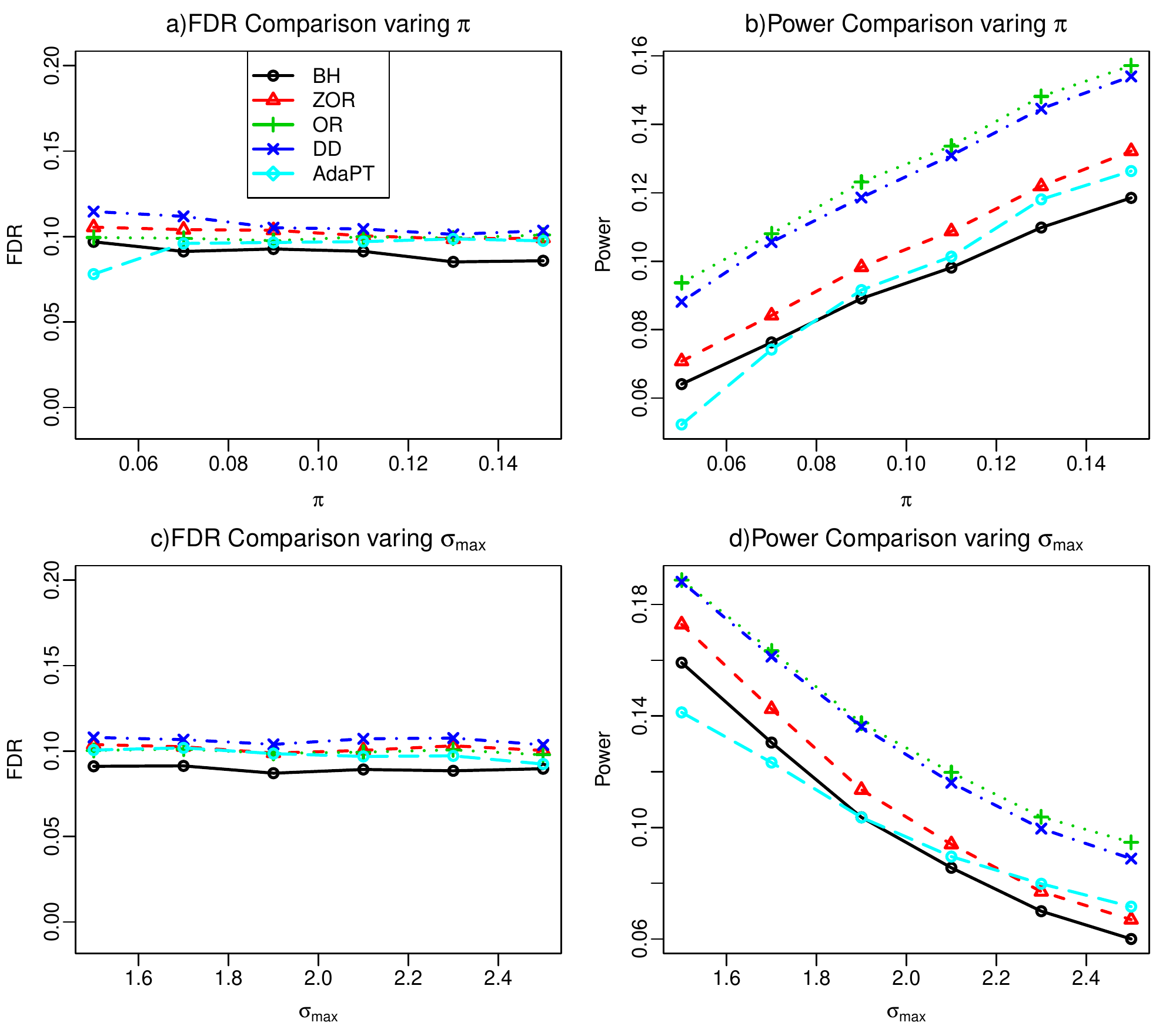}
		\caption {\label{genearlmu} Comparison when $\sigma_i$ is generated from  a uniform distribution and $\mu_i$ is generated from a Gaussian mixture. We vary $\pi$ in the top row and  $\sigma_{max}$ in the bottom row. All methods control the FDR at the nominal level. DD performs almost as well as OR and has a significant power advantage over ZOR,BH and AdaPT.}
	\end{center}
\end{figure}

We generate $\sigma_i$ uniformly from $[0.5, \sigma_{max}]$, and generate $X_i$ according to the following model:
$$X_i|\sigma_i\overset{iid}{\sim} (1-\pi)N(0, \sigma_i^2)+\pi N(\mu_i, \sigma_i^2),\ \ \ \mu_i\overset{iid}{\sim}0.5N(-1.5,0.1^2)+0.5N(2,0.1^2).$$
 In the first setting, we fix $\sigma_{max}=2$ and vary $\pi$ from 0.05 to 0.15. In the second setting, we fix $\pi=0.1$ and vary $\sigma_{max}$ from 1.5 to 2.5. {Five} methods are compared: the ideal full data oracle procedure (OR), the $z$-value oracle procedure of \citep{SunCai07} (ZOR), the Benjamini-Hochberg procedure (BH), AdaPT \citep{LeiFit18} (AdaPT), and the proposed data--driven HART procedure (DD). The nominal FDR level is set to $\alpha=0.1$. For each setting, the number of tests is $m = 20,000$. Each simulation is also run over $100$ repetitions. The results are summarized in Figure \ref{genearlmu}.
All methods can control the FDR at the nominal level with BH slightly conservative. DD performs almost as well as OR. The ordering information from $\sigma_i$ seems to help AdaPT in some cases, but in other cases, it causes AdaPT to underperform BH. There is a clear power gap between DD and ZOR.

\subsection{Unknown $\sigma_i$}
\begin{figure}[t!]
	\begin{center}
		\includegraphics[width=5in]{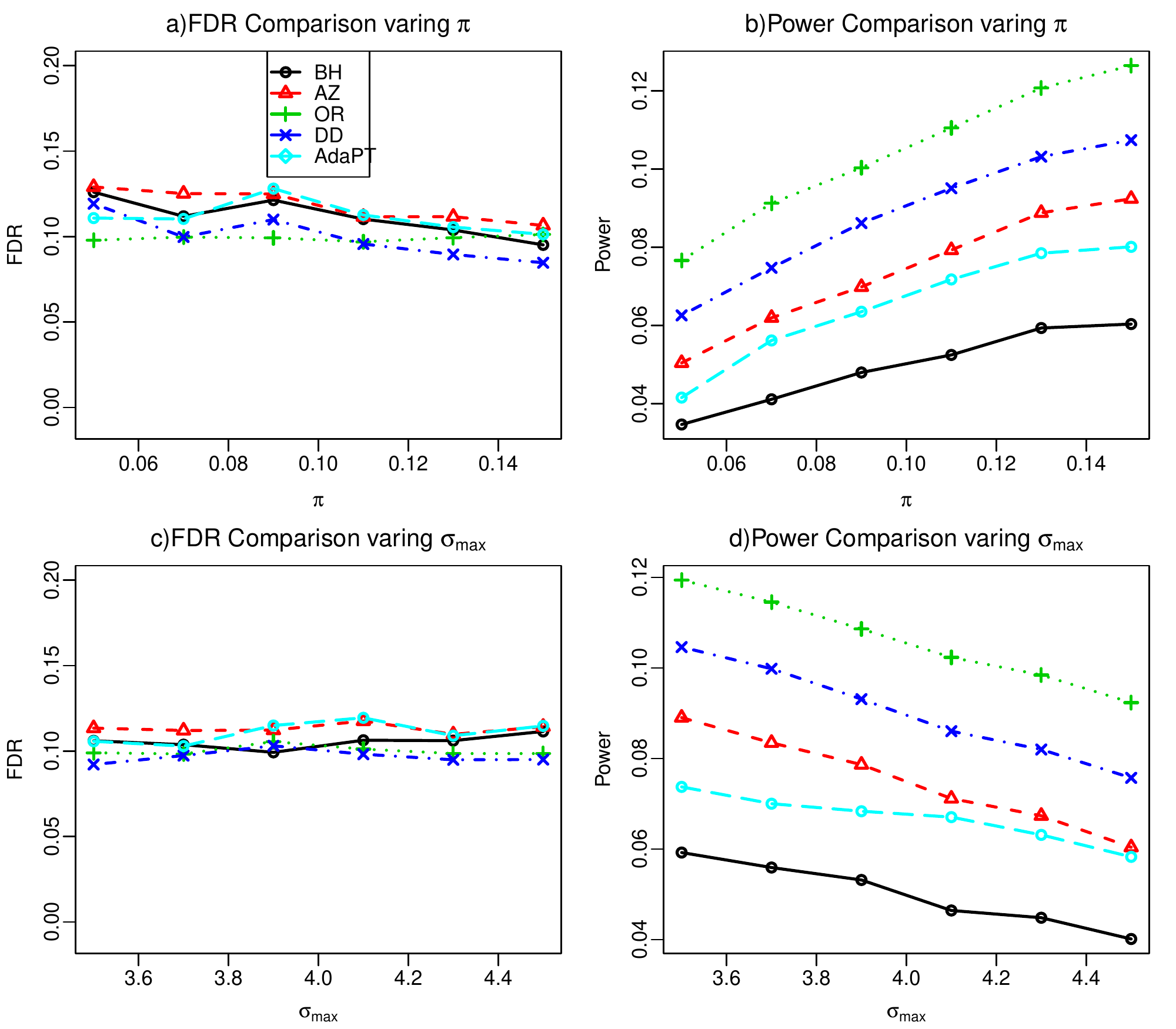}
		\caption {\label{estsig} Comparison when $\sigma_i$ is unknown. We vary $\pi$ in the top row and  $\sigma_{max}$ in the bottom row. All data-driven methods are adversely affected. But the effect on DD is the smallest.}
	\end{center}
\end{figure}

This section investigates the robustness of our method when $\sigma_i$ is unknown.
In some applications, the exact value of $\sigma_i$ is unknown but can be estimated.  For this simulation, we independently generate $200$ copies of $X_i$ using the following model:
$$X_i|\sigma_i\overset{iid}{\sim} (1-\pi)N(0, \sigma_i^2)+\pi N(2/\sqrt{200}, \sigma_i^2),\ \ \ \sigma_i\overset{iid}{\sim}U[0.5,\sigma_{max}].$$ For fair comparison we replace ZOR by AZ, the data driven version of ZOR described in \citep{SunCai07}. We use the sample standard deviation of $x_i$ (denoted  $s_i$) as an estimate of $\sigma_i$ for DD, AZ, BH and AdaPT. OR has access to the exact value of $\sigma_i$. We then apply the testing procedures to the pairs $(\sqrt{200}\bar{x}_i, s_i)$. The z-value is computed as $\dfrac{\sqrt{200}\bar{x}_i}{s_i}$ and the p-value is computed using $\frac{1}{2}\{1-\Phi(|z_i|)\}$ where $\Phi$ is the CDF for standard normal distribution. Strictly speaking the z-values should follow a $t$ distribution, but since the sample size is relatively large, normal distribution serves as a good approximation.  The number of tests is $m = 20,000$. Each simulation is run over $100$ repetitions.  In the first setting, we fix $\sigma_{max}=4$ and vary $\pi$ from 0.05 to 0.15. In the second setting, we fix $\pi=0.1$ and vary $\sigma_{max}$ from 3.5 to 4.5. The results are summarized in Figure \ref{estsig}.

We can see that all data-driven methods have been adversely affected. However, the inflation of the FDR for DD is less severe compared to the other three data-driven methods. The gap in power between OR and DD becomes larger but the power advantage of DD over other methods is maintained.

\subsection{Weak dependence}
\begin{figure}[t!]
	\begin{center}
		\includegraphics[width=5in]{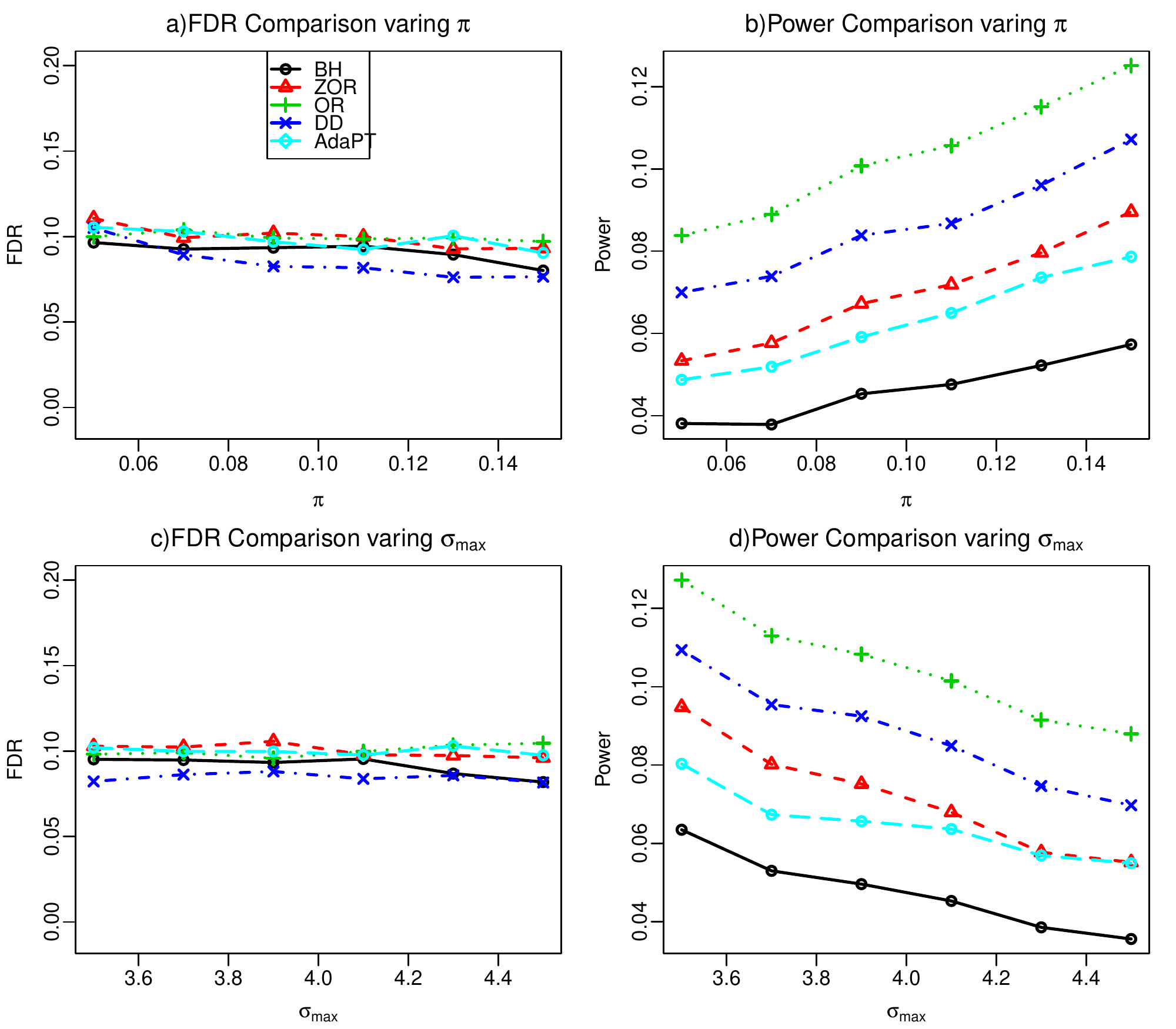}
		\caption {\label{dep1} Comparison under weak dependence case 1. The dependence structure has little effect on all the methods. The pattern is almost the same as in the independent case.}
		\label{plot:dep1}
	\end{center}
\end{figure}

\begin{figure}[t!]
	\begin{center}
		\includegraphics[width=5in]{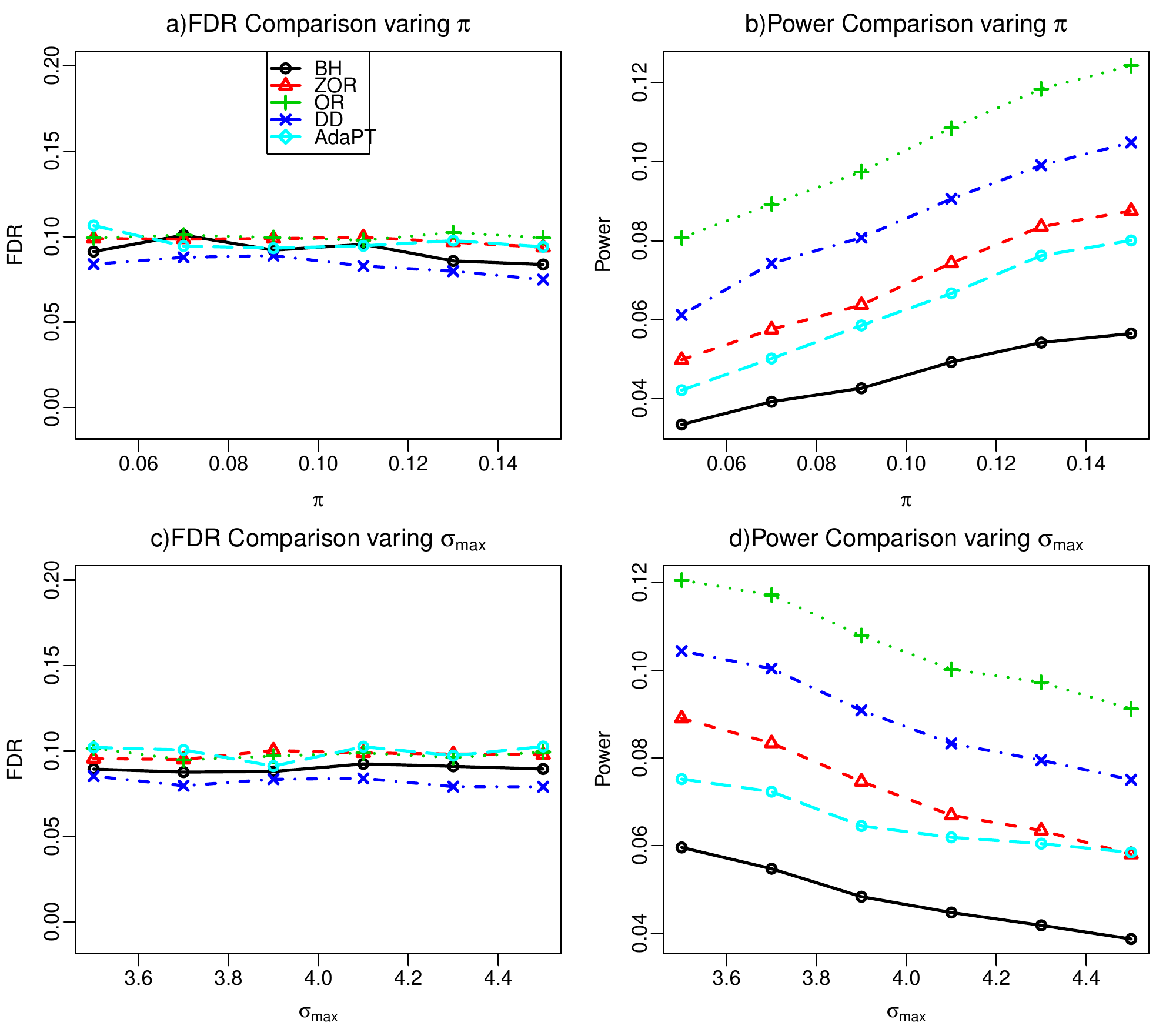}
		\caption {\label{dep2} Comparison under weak dependence case 2. Again, the dependence structure has little effect on all the methods. The pattern is almost the same as in the independent case.}
	\end{center}
\end{figure}

We investigate the robustness of our method under weak dependence. We consider two weak dependence models:\\
\textit{Case 1}: We use the following model:
$$\mu_i \overset{iid}{\sim} (1-\pi)\delta_0(\cdot)+\pi\delta_{2}(\cdot),\quad \sigma_i \overset{iid}{\sim} U[0,\sigma_{max}],\quad \pmb{x}\sim N(\pmb{\mu},\Sigma),$$
where the covariance matrix is a block matrix:
\[ \Sigma = \begin{pmatrix}
M_{11} & 0\\\
0 & M_{22}
\end{pmatrix}.
\]
Here $M_{11}$ is a $4000\times 4000$ matrix, the $(i,i)$ entry is $\sigma^2_i$, the $(i,i+1)$ and $(i+1,i)$ entries are $0.5\sigma_i\sigma_{i+1}$, the $(i,i+2)$ and $(i+2,i)$ entries are $0.4\sigma_i\sigma_{i+2}$. The rest of the entries are 0. $M_{22}$ is a $16000\times 16000$ diagonal matrix, with the $(i,i)$ entry being $\sigma^2_{i+4000}$. In the first setting, we fix $\sigma_{max}=4$ and vary $\pi$ from 0.05 to 0.15. In the second setting, we fix $\pi=0.1$ and vary $\sigma_{max}$ from 3.5 to 4.5. The results are summarized in Figure \ref{dep1}. The number of tests is $m=20,000$.
We can see that the weak dependence has little effect on the pattern. This demonstrates the robustness of DD.\\
\textit{Case 2}: We use the following model:
$$\mu_i \overset{iid}{\sim} (1-\pi)\delta_0(\cdot)+\pi\delta_{2}(\cdot),\quad \sigma_i \overset{iid}{\sim} U[0,\sigma_{max}],\quad \pmb{x}\sim N(\pmb{\mu},\Sigma),$$
where $\Sigma = \begin{pmatrix}
M_{11} & 0\\\
0 & M_{22}
\end{pmatrix}.
$
Here $M_{11}$ is a $4000\times 4000$ matrix, the $(i,j)$ entry of $M_{11}$ is $0.5^{|i-j|}\sigma_i\sigma_j$.
$M_{22}$ is a $16000\times 16000$ diagonal matrix with diagonal being $(\sigma^2_{4001},...,\sigma^2_{20000})$. In the first setting, we fix $\sigma_{max}=4$ and vary $\pi$ from 0.05 to 0.15. In the second setting, we fix $\pi=0.1$ and vary $\sigma_{max}$ from 3.5 to 4.5. The results are summarized in Figure \ref{dep2}.
We can see again, there is no noticeable difference in pattern from the independent setting. This shows the robustness of DD under positive dependence. This seems to be consistent with existing results in the literature.

\subsection{Non-Gaussian noise}
\begin{figure}[t!]
	\begin{center}
		\includegraphics[width=5in]{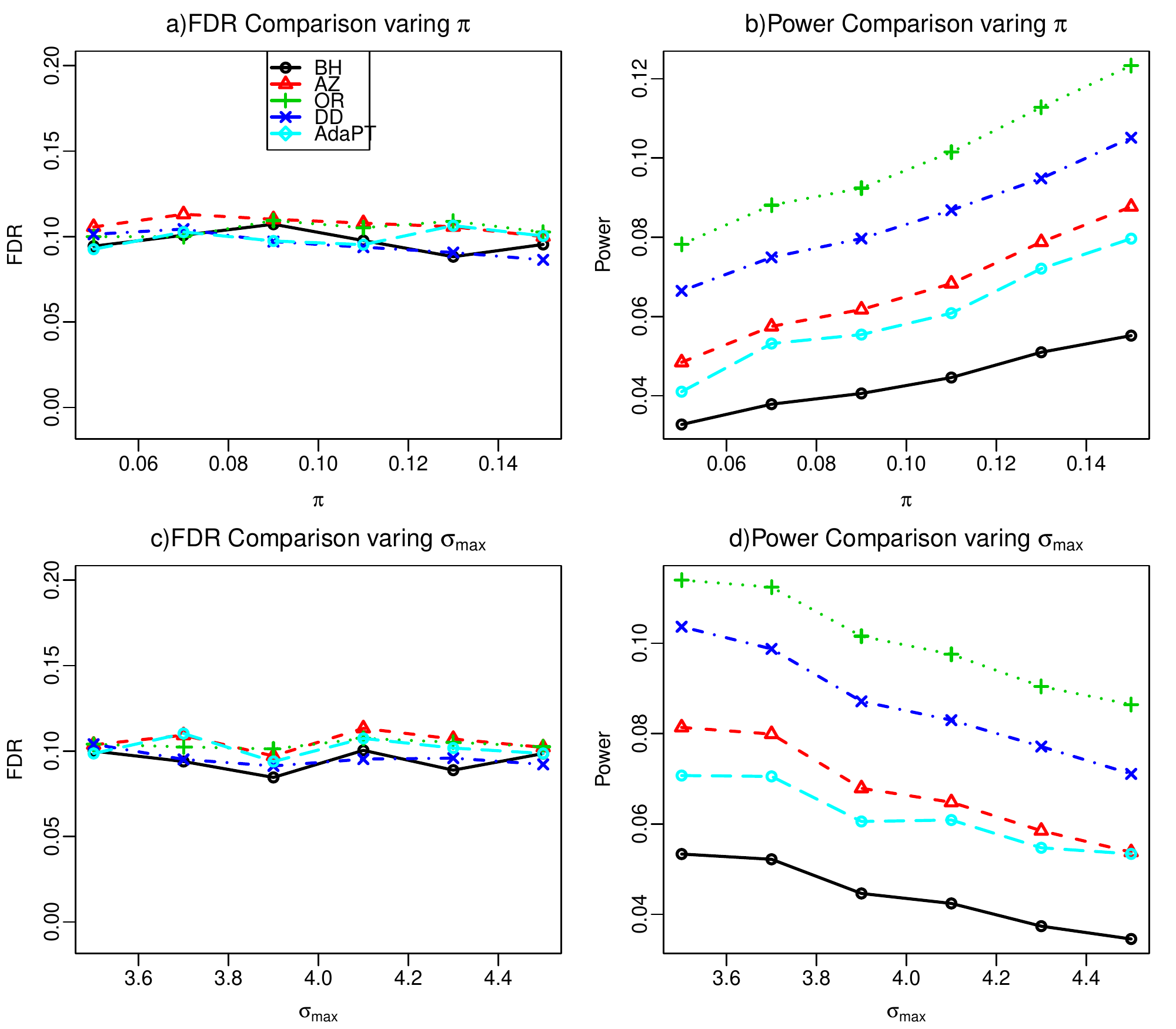}
		\caption {\label{ng} Comparison when the noise is non-Gaussian. The non-Gaussian noise has little effect on the overall pattern.}
	\end{center}
\end{figure}

We study the performance of our method when the noise follows a heavy-tailed distribution. For this simulation, we independently generate $200$ copies of $X_i$ using the following model:
$$X_i=\mu_i+\sigma_i\epsilon_i,\quad \mu_i \overset{iid}{\sim} (1-\pi)\delta_0(\cdot)+\pi\delta_{2/\sqrt{200}}(\cdot),\quad \sigma_i \overset{iid}{\sim} U[0,\sigma_{max}],\quad \epsilon_i\overset{iid}{\sim}t_5.$$
 We use the sample standard deviation of $x_i$ (denoted $\hat{\sigma}_i$) as an estimate of $\sigma_i$ for all five methods. We then apply the testing procedures to the pairs $(\sqrt{200}\bar{x}_i,\hat{\sigma_i})$. The number of tests is $m = 20,000$. Each simulation is run over $100$ repetitions. Since the precise distribution of $x_i/\hat{\sigma}_i$ is hard to compute, we replace ZOR by AZ.
 Note that in this case, the model is mis-specified even for OR. But OR has access to the distribution of $\mu_i$ and $\pi$ while other data-driven methods do not.
 In the first setting, we fix $\sigma_{max}=4$ and vary $\pi$ from 0.05 to 0.15. In the second setting, we fix $\pi=0.1$ and vary $\sigma_{max}$ from 3.5 to 4.5. The results are summarized in Figure \ref{ng}. We can see that the pattern is similar to the Gaussian-noise case.

\subsection{Unknown null distribution}
\begin{figure}[ht!]
	\begin{center}
		\includegraphics[width=5in]{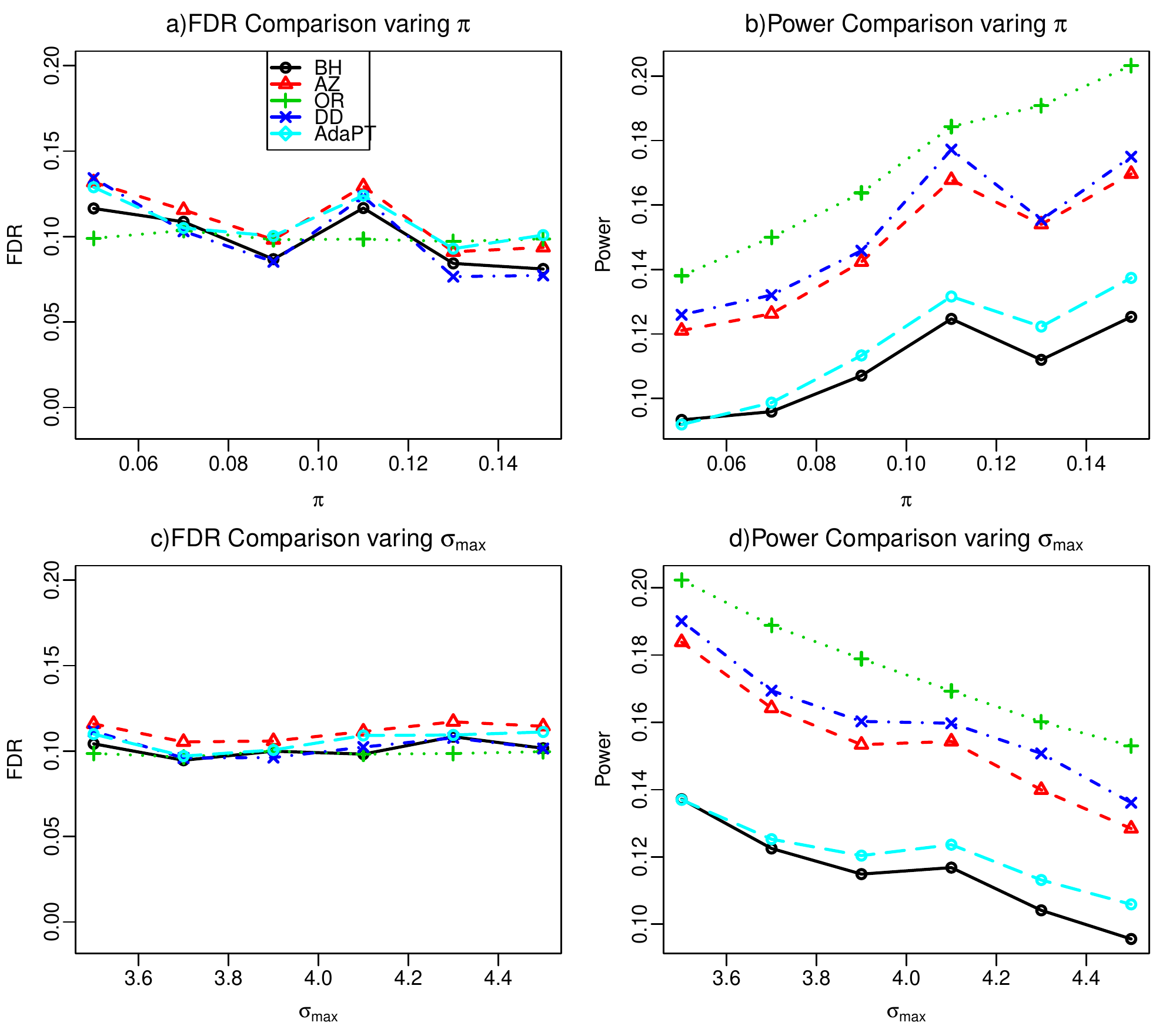}
		\caption {\label{emp} Comparison when the null distribution is estimated. We can see that all the data-driven methods show some instability: this is due to the variance of the estimator of null variance. But DD still out-performs other data-driven methods  }
	\end{center}
\end{figure}

We study the performance of our method when the z-values do not follow a standard normal distribution under the null hypothesis. For this simulation, we use the following model:
$$Z_i \overset{iid}{\sim} N(0,0.8^2),\quad \sigma_i \overset{iid}{\sim} U[0,\sigma_{max}],\quad X_i=Z_i\sigma_i+\mu_i,\quad  \mu_i \overset{iid}{\sim} (1-\pi)\delta_0(\cdot)+\pi\delta_{2}(\cdot).$$
For the data-driven methods, we estimate the null distribution of $z_i$'s using the method described in \cite{JinCai07}. Let $\sigma_0$ be the estimated variance of $Z_i$. For DD, $f_{0,\sigma_j}$ is now the density function of $N(0,\sigma^2_0\sigma^2_j)$. In the first setting, we fix $\sigma_{max}=4$ and vary $\pi$ from 0.05 to 0.15. In the second setting, we fix $\pi=0.1$ and vary $\sigma_{max}$ from 3.5 to 4.5. The results are summarized in Figure \ref{emp}.

We can see that the variance of the estimator of null variance has a noticeable effect on all data-driven methods. But DD still performs the best.

\end{document}